\documentclass[twocolumn]{aastex7}
\usepackage{graphicx}	
\usepackage{amsmath}	
\usepackage{amssymb}
\usepackage{tikz}
\usepackage{placeins}
\usepackage{afterpage}
\usepackage{soul}

\usepackage{newtxtext,newtxmath}

\shorttitle{Simulating TRGs}
\shortauthors{Mondal et al.}
\graphicspath{{./}{Figures/}}

\begin{document}

\title{Low-velocity precessing jets can explain observed morphologies in the Twin Radio Galaxy TRG\,J104454+354055
}

\author{Santanu Mondal}
\affiliation{Indian Institute of Astrophysics, II Block Koramangala, Bangalore 560034, India}
\email[show]{santanuicsp@gmail.com}

\author{Gourab Giri}
\affiliation{Department of Physics, University of Pretoria, Private Bag X20, Hatfield 0028, South Africa}
\affiliation{South African Radio Astronomy Observatory, 2 Fir St, Black River Park, Observatory 7925, South Africa}
\email[]{gourab.giri@up.ac.za }

\author{Ravi Joshi}
\affiliation{Indian Institute of Astrophysics, II Block Koramangala, Bangalore 560034, India}
\email[]{ravi.joshi@iiap.res.in}

\author{Paul J.\ Wiita}
\affiliation{Department of Physics, The College of New Jersey, 2000 Pennington Road, Ewing, NJ 08628-0718, USA}
\email[]{wiitap@tcnj.edu }

\author{Gopal-Krishna}
\affiliation{UM-DAE Centre for Excellence in Basic Sciences, Mumbai, 400098, India}
\email[]{gopaltani@gmail.com }

\author{Luis C.\ Ho}
\affiliation{Kavli Institute for Astronomy and Astrophysics, Peking University, Beijing, 100871, China}
\affiliation{Department of Astronomy, School of Physics, Peking University, Beijing, 100871, China}
\email[]{lho.pku@gmail.com }

\begin{abstract}
Our understanding of large-scale radio jets in merger systems has been drastically improved in the era of VLA, VLBA/EVN, uGMRT, and MeerKAT. Twin Radio Galaxies (TRGs) are the rare interacting galaxy pairs where both supermassive black holes host kiloparsec-scale bipolar radio jets. Only recently was a third TRG discovered and it shows significantly different jet morphologies than the previous two. Due to both the extreme paucity and complexity of such systems, the launching of their jets as well as their mutual interaction during the propagation through the ambient medium are not well understood. We have performed 3D hydrodynamic simulations to study the bipolar jets in the third TRG, J104454+354055. Our study indicates that the precession of mutually tilted bipolar jets originating from the two galactic nuclei separated by tens of kiloparsecs and propagating at low velocities can explain the observed morphologies. The simulated jet precession timescales are short compared to the overall dynamical timescale of the jets and could originate from Lense-Thirring effects in the accretion disks. This approach to understanding the TRG jet dynamics could also be applied to other TRG systems with similar helical morphologies that may be discovered in the upcoming era of the SKA and its pathfinder surveys.
\end{abstract}

\keywords{galaxies: clusters: general -- galaxies: active -- galaxies: jets -- jets:ISM -- quasars: supermassive black holes -- method: numericals
}



\section{Introduction} \label{sec:intro}

Most galaxies are believed to host a supermassive black hole (SMBH) at the center \citep{KormendyRichstone1995ARA&A..33..581K,MagorrianEtal1998AJ....115.2285M}, whose evolution is primarily thought to result from a sequence of galaxy mergers. These mergers channel large amounts of fresh matter towards the center, triggering an accelerated accretion process \citep{Fragile2007,Hopkins2012}. However, the complex interplay between matter accretion, SMBH growth, and matter ejection or outflows, both thermal and nonthermal, remains a key subject of frontier research \citep[e.g.][]{Gaspari2020}.

Following the collision of two galaxies, it is expected that the SMBHs in each host will gradually sink toward each other in their orbits due to stellar dynamical friction and/or energy dissipation caused by the circumnuclear gas, ultimately forming a binary or a dual SMBH system \citep[][and references therein]{BegelmanEtal1980Natur.287..307B,KormendyHo2013ARA&A..51..511K}. 
When the approaching active SMBHs are separated by relatively small ($<100$ pc) distances they are referred to as binary active galactic nuclei \citep[BAGN; see][]{RodriguezEtal2006ApJ...646...49R,BansalEtal2017ApJ...843...14B,kharb.lal.etal.2017}, and when still separated by kiloparsec-scales they are called dual AGN (DAGN) \citep[e.g.][]{Burke-Spolaor2014arXiv1402.0548B,RubinurEtal2018JApA...39....8R}. 

Although several detections of DAGN have been made, identifying binary systems remains challenging due to the requirement of very high-resolution observations across multiple frequency bands \citep[see,][]{DeRosaEtalReview2019NewAR..8601525D,BreidingEtal2022ApJ...933..143B}. Some candidates are even suspected to host BAGNs beyond the resolution capabilities of most contemporary telescopes \citep{Kharb2019}. Furthermore, different observational features, such as double-peaked optical spectra or curved morphologies of double radio lobes, once considered to be signatures for identifying such binary candidates, have been found to be explicable in terms of other astrophysical processes occurring around the violent nuclear environment  \citep[e.g.,][]{Fu2012, Rubinur2019}. Regardless of these potential alternatives, the indirect signatures of DAGNs, such as S- or X-shaped radio galaxies \citep[e.g.][]{BegelmanEtal1980Natur.287..307B,Rottmann2001PhDT.......173R,GoaplEtal2003ApJ...594L.103G,NandiEtal2017MNRAS.467L..56N} or optical periodicity \citep[e.g.][]{GrahamEtal2015Natur.518...74G,LiuEtal2015ApJ...803L..16L}, continue to play a crucial role in identifying DAGN candidates among the vast population of jetted AGNs.
This set of systems provide a crucial step towards understanding the final stages of galaxy mergers (known as the final parsec problem), when SMBHs are separated by (sub)parsec scales and can emit gravitational waves contributing significantly to the gravitational wave background \citep[e.g.][]{agazie.etal.2023}. Several multi-wavelength surveys are now focusing on detecting such merger systems, with an emphasis on gravitational-wave detection as an emerging branch of astronomy \citep[][and references therein]{AbazajianSDSS..Survey2009ApJS..182..543A,WrightWISE..Survey2010AJ....140.1868W,AhnBOSS..SDSS..Survey2012ApJS..203...21A}.

Depending on the quantity of gas that becomes available for accretion during a galaxy merger (wet versus dry merger), either one or both SMBHs in the binary orbit may launch relativistic plasma jets. While jet ejection from a single massive AGN is relatively common \citep{Murgia2001,Bruno2019}, simultaneous jet ejection from both AGNs is a rare phenomenon. Such a rare system is labeled as a Twin Radio Galaxy (TRG) in the literature. 
The first case of such a dual jet-pair ejection was discovered 
four decades ago by \citet{OwenEtal1985ApJ...294L..85O}: 3C\,75 near the center of the cluster Abell 400 ($z = 0.023$). It is a Wide-Angle-Tail (WAT) radio source with overall size of $\sim0.5$ Mpc. Recall that the spectacular WAT morphology observed in 3C 465 has been attributed to the high-velocity orbital motion of the jetted AGN about another massive galaxy with which it is merging \citep{WirthEtAl1982AJ.....87..602W}. The second TRG, which was discovered three decades ago, is PKS\,2149-158, of size $\sim 0.4$ Mpc, located near the outskirts of Abell 2382 ($z = 0.062$) \citep{ParmaEtal1991AJ....102.1960P,GuidettiEtal2008A&A...483..699G}. Only recently, the third TRG, J104454+354055 (hereafter TRG\,J104454), with  $z = 0.162$, was discovered by \citet{Gopal-KrishnaEtal2022MNRAS.514L..36G} using data from the upgraded Giant Metrewave Radio Telescope (uGMRT).  In this TRG, both radio jets are discernible up to $\sim0.1$ Mpc scale and the overall radio structure spans $\sim0.4$ Mpc with Fanaroff-Riley class I (FR I) morphology.  Initially, this TRG had been identified as a candidate for an X-shaped radio galaxy \citep{YangEtal2019PhRvL.123r1101Y,JoshiEtal2019ApJ...887..266J}, but later it was classified as a TRG. 

Note that possible evidence for DAGNs, with both galactic nuclei capable of ejecting jets, was also speculated from the anomalous spectral behavior observed in some X-shaped (winged) radio galaxies \citep{Lal2005,Lal2019}. However, recent studies have proposed alternative explanations for this puzzling behavior or have failed to confirm the reported spectral anomaly using radio maps with a wider radio-frequency coverage, and this underscores the need for further investigation into the general applicability of such models \citep{GiriEtal2022A&A...662A...5G,Patra2023}.

Given just the handful of TRGs discovered so far, much remains uncertain about these complex systems. 
Some open questions include: the role of persistent accretion and SMBH dynamical and physical parameters on the jet ejections and their evolution; the relative stability of the gaseous haloes of the interacting galaxies; and the impact of multiple radio jets in (de)stabilizing the surrounding environment.

To investigate the environmental influences on bipolar-jetted systems and also to understand the jet evolution phases of such systems, here we initiate a program of numerical modeling of the recently discovered source TRG J104454. Similar simulations have been conducted for the TRG 3C 75 \citep{MolnarEtal2017ApJ...835...57M,Musoke2020}. In comparison to the previously known TRGs, the configuration of the TRG J104454 is distinctly neater since it does not appear to have any significant intracluster cross-wind \citep{Gopal-KrishnaEtal2022MNRAS.514L..36G}.  Note that the jets in the previous two TRGs eventually became C-shaped, most likely due to the effect of such cross-winds \citep{OwenEtal1985ApJ...294L..85O,GuidettiEtal2008A&A...483..699G}. Models have demonstrated that a variety of observed jet morphologies in an individual or twin galaxies can arise either from precession effects in jets \citep{GowerEtAl1982ApJ...262..478G} or gravitational interactions of the merging galaxies which can affect the jets \citep{WirthEtAl1982AJ.....87..602W}. The TRG J104454 is likely exhibiting jet precession and potential lobe collisions at larger-scales, but with an apparent lack of environmental cross-wind and consequent distortions of the lobe structures. Therefore, it is overall a simpler testbed for studying the nature of DAGNs as traced using their larger-scale radio structures. Regardless of the mechanism responsible for the precession of the jets, we performed simulations of the TRG under the premise of precessing jets and compared their outcomes with the observed uGMRT morphology of the present TRG.

In the next section, we discuss the formulation of our simulations. In \S 3, the simulation results are discussed in terms of interpreting the properties recovered from observations of TRG  J104454 including the possible origin of the inferred jet precession. We summarize our conclusions in \S 4.

\section{Simulated Configuration}
\begin{figure}
\centering
\includegraphics[height=6.0truecm,width=8.0truecm,angle=0,trim={4.0cm 1.0cm 5.0cm 0.0cm}, clip]{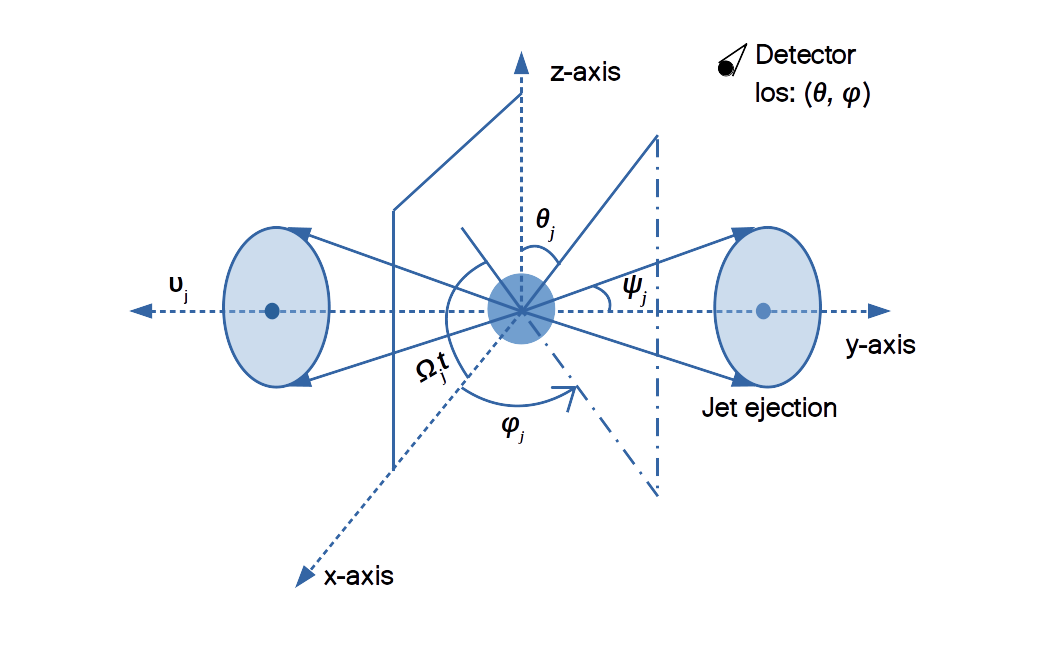}
\caption{3D illustration of the injection of a bipolar jet \citep[similar to][]{GiriEtal2022A&A...662A...5G} based on our simulation configuration. The bipolar jet precesses around the y-axis at an angle $\Psi_j$ with a precession period $P_j$, resulting in an angular velocity $\Omega_j = 2 \pi/P_j$ in the x-z plane. Also sketched here is the line of sight
position $(\theta, \varphi)$, along which the mock emission maps were generated.
}
\label{fig:Illustris}
\end{figure}

The bound pair of radio galaxies `TRG J104454' measures $\sim 220$ kpc and beyond, along the bipolar jet-flow axis and has been identified as a low-powered jet-driven system \citep{Gopal-KrishnaEtal2022MNRAS.514L..36G}.  To simulate such an extended configuration, we employed the hydrodynamical module of the PLUTO code in a three-dimensional Cartesian system \citep{Mignone2007}. An  illustration of the simulation configuration of a single bipolar jet is shown in \autoref{fig:Illustris}, where both the injection and precession of the jet are displayed. We note that the jet nozzles (through which the jet plasma is being injected) have not been assigned a specified opening angle, as we expect the jet structures observed at large scales to be governed by the overall dynamics and precession of the jets \citep{GiriEtal2022MNRAS.514.5625G}. 
As discussed below, we choose units for the three basic parameters to be: length ($l_0$) as 10 kpc; velocity ($\varv_0$) as $6.9\times10^7$ cm s$^{-1}$ (sound speed in the
jet); and density ($\rho_0$) as 0.03 amu cm$^{-3}$ (at the center of the ambient gaseous halo).
All other relevant parameters can be scaled based on these parameters, e.g., the simulation time can be scaled as (unit length/unit
velocity) which is $\approx14$ Myr and the kinematic luminosity ($L_{kin0}=\rho_0 l_0^2 \varv_0^3$) can be scaled as $1.5\times10^{43}$ erg s$^{-1}$. 

In order to describe the ambient medium through which the jets propagate, we use a spherical King density profile to mimic a rich galaxy group environment \citep[e.g.][]{Hardcastle2013}. This relies on the observation that the twin jets extend far beyond the stellar boundaries of their host galaxies and evolve in a dynamically active galaxy group \citep{Gopal-KrishnaEtal2022MNRAS.514L..36G}. This premise also holds true for the other two TRGs reported to date, although their radio structures show additional complicating features related to cluster cross-winds (Sec. 1). The gas density distribution is defined as

\begin{equation}
    \rho(r)=\rho_0 \left[1+\left(\frac{r}{r_c}\right)^2 \right]^{-3\beta/2}~~,
\end{equation}
where $r$ is the radial distance from the center of the computing domain (also the center of the ambient medium) at \{0, 0, 0\}, taken as midway between the origins of the two injected bipolar jets and is evaluated as $(x^2+y^2+z^2)^{1/2}$; $\rho_0$ is the (plasma) gas density of the ambient cluster medium at $r = 0$ (adopted to be 0.03 amu cm$^{-3}$); $r_c$ is the core radius, taken to be $125$ kpc, and $\beta$ is $0.55$ \citep[e.g.,][]{Hardcastle2018,Giri2023}. We adopted an ideal equation of state for the ambient gas and applied a constant pressure profile (of magnitude $2.5 \times 10^{-12} \, {\rm dyn ~cm}^{-2}$), in order to establish a static equilibrium of the ambient medium as the initial condition, while the jets are being injected. These are typical parameter sets adopted for representing the large-scale jet environment \citep[e.g.,][]{Musoke2020}. The adopted sound speed in each jet, $\varv_0$ = $6.9\times10^7$ cm s$^{-1}$, corresponds to a jet temperature of $\sim 2.2$ keV \citep{Kharb2019}. Then a jet Mach number of  $M_j$ = 65.2 is adopted as a default parameter for the injected jets, which is equivalent to a jet velocity, $\varv_j = 0.15c$ (selected to mimic a low jet power consistent with the estimates reported for TRG J104454); $c$ is the speed of light. The simulation domain extends from $-70$ kpc to $+70$ kpc in the $x$- and $z$-directions and $-150$ kpc to $+150$ kpc in the $y$-direction, with uniformly distributed \{$336\times720\times336$\}  grids. We employ outflow boundary conditions for the boundaries of the computational domain.

We inject two non-parallel bipolar jets with their origins separated by 30 kpc \citep{Gopal-KrishnaEtal2022MNRAS.514L..36G} from the center of the ambient medium configuration (also taken as the center of the simulation). Both pairs of jets are injected through cylindrical injection zones, which are divided into three launch regions: the top region (producing the flow of the jet primarily along the $+y$ axis); the bottom, (for the flow mostly along the $-y$ axis); and a neutral region (sandwiched between the two zones, where none of the parameters vary) around the center \citep[similar to][]{RossiEtal2017A&A...606A..57R} with injection radius ($R_j$) usually set to 4 kpc and the height of each cylinder ($H_j$) being 5 kpc. This relatively large value of $R_j$ is necessary to resolve the jets with sufficient grid points and to generate enough mechanical power for the jet to overcome the resistance due to the environment. For the grid resolution mentioned above, the jet launching diameter  contains around 10 grid cells (a convergence test has also been performed using a higher resolution; see Appendix A.)

In our simulations we continuously inject bipolar plasma jets from the cylindrical nozzles, with velocities $\varv_{r}$ and $\varv_{l}$. Here, the subscripts $r$ and $l$ denote left and right jets respectively, as each injected jet is bi-polar. Guided by the observed morphologies, we have tilted each jet at respective angles $\theta_{r}$ and $\theta_{l}$ (from the +$y$ axis), with precession cone half-angles of $\Psi_{r,l}$ and periods $P_{r}$ and $P_{l}$. We followed the geometric model proposed by \citet{HjellmingJohnston1981ApJ...246L.141H} to implement the precession. Accordingly, both jets are tilted with respect to the $y$-axis and their precession axes are parallel to the $y$-axis. Therefore, the velocity components along all three axes of the precessing, tilted jets can be written as

\begin{equation}
    \begin{split}
        \varv_{j, x} &= \varv_{j} \sin \Psi_{j} \cos (\Omega_{j}t) \sin \theta_{j},\\
        \varv_{j, y} &= \varv_{j} \cos \Psi_{j} \cos \theta_{j}, ~~{\rm and}\\
        \varv_{j, z} &= \varv_{j} \sin \Psi_{j} \sin (\Omega_{j}t),
    \end{split}
\end{equation}
where $\Omega_{j} (=2\pi/P_{j})$ is the precession frequency for each jet and the subscript $`j'$ denotes the quantities for the left $(l)$ and right $(r)$ bipolar jets.  The kinetic power of both these sub-relativistic jets is estimated using the relation \citep{HardcastleHrause2014MNRAS.443.1482H}

\begin{equation}
    L_{\rm kin}=\pi R_{j}^2 M_{j}\left(\frac{1}{2}M_{j}^2\rho_j+\frac{5}{2}T_j\rho_j\right),
\end{equation}
with $T_j$ the jet plasma's temperature.

Using the initial jet parameters, the estimated $L_{\rm kin}$ we considered ranges from $8.8\times10^{44}$ erg s$^{-1}$ for the minimum jet velocity, which was for $M_j=60.9$, to $2.6\times10^{45}$ erg s$^{-1}$ for the maximum jet velocity examined, corresponding to $M_j=87$. The parameters adopted for the selected TRG simulations, which are displayed in the following figures, are listed in \autoref{table:RunPars}. Note that the jet parameters, with the exception of $\Psi_j$, are taken independently for individual bipolar jets during each simulation.  
 Implementing the above processes and grid resolutions, we have checked how well the results converge for the setup S6. All results are presented for the grid distribution of $\{336 \times 720 \times 336\}$ taking into account the convergence of results with higher resolution  (see Appendix \ref{sec:appendA}). Also, all snapshots are generated for a tracer that tracks the presence of jet material in a grid zone, set at a threshold of 0.5.

\begin{table}
\caption{\label{table:RunPars} Input parameters adopted for the selected TRG simulations presented in \autoref{fig:SimSetup1} and \autoref{fig:Setup2to6}. }
\hspace{-1.8cm}
\begin{tabular}{ccccccccc}
\hline
Setup & $\theta_l$ & $\theta_r$ &$P_{r}$ &$P_{l}$ &$\Psi_{j}$ &$M_{j}$&$M_{j}$&$R_j$\\
&(Deg)&(Deg)&(Myr)&(Myr)&(Deg)&(right)&(left)&(kpc)\\
\hline
S1&-45&40&49&28&20&65.2&65.2&4\\
S2&-45&-40&49&28&20&65.2&65.2&4\\
S3&-45&25&28&28&20&65.2&65.2&4\\
S4&-40&45&49&49&30&82.6&65.2&4\\
S5&40&45&35&35&30&87.0&60.9&4\\
S6&-45&-30&28&28&23&65.2&65.2&4\\
S7&-45&40&49&28&20&65.2&65.2&2\\\hline
\end{tabular} 

Notes: For each simulation $\theta$ is the tilt angle of the bipolar jet at injection point, relative to the y-axis, $P$ is the precession period of the jet, $\Psi_j$ is the precession angle, $M_j$ denotes the Mach number of the jet and $R_j$ is the jet radius. We note that all of our simulations are carried in dimensionless quantities, and then converted back into physical quantities by scaling with the respective units, adopted and derived.
\end{table}

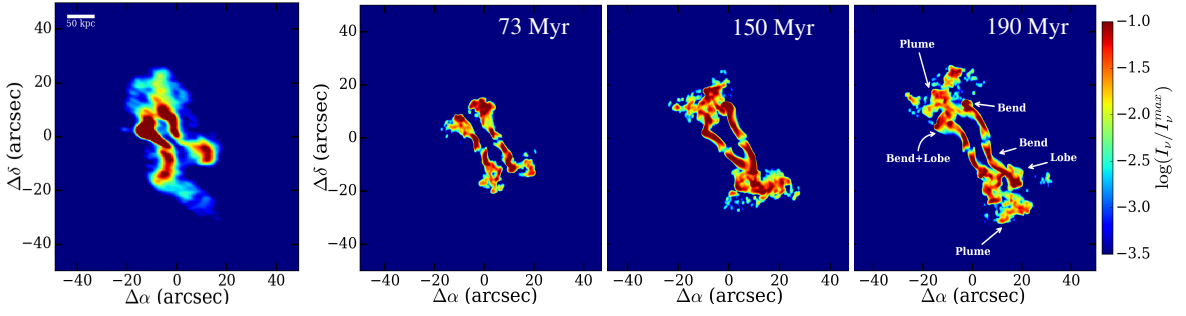
\begin{figure*}
\centering
\begin{tikzpicture}
\draw (0, 0) node[inner sep=0] {\raisebox{0.1cm}
{\includegraphics[height=4.0truecm,width=4.8truecm,angle=0,trim={0.0cm 0.0cm 2.5cm 1.5cm}, clip]{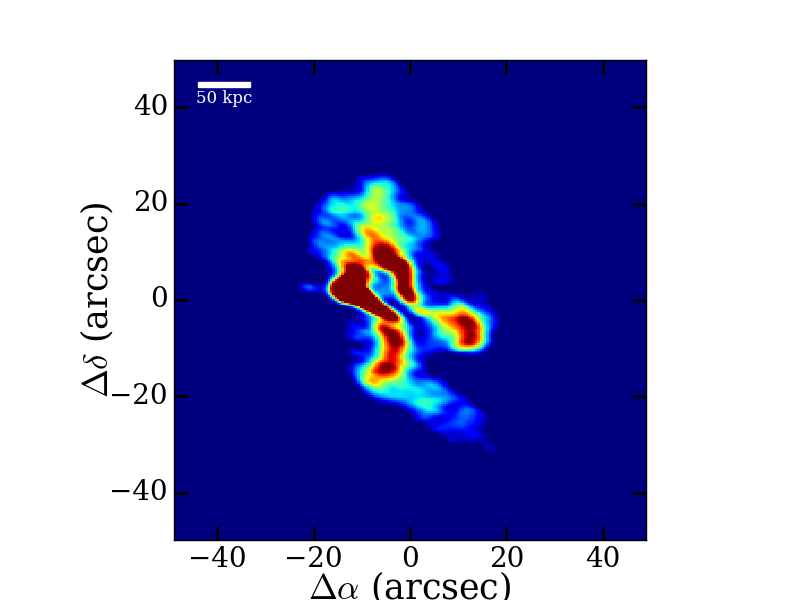}}};
\end{tikzpicture}
\hspace{-0.5cm}
\begin{tikzpicture}
\draw (0, 0) node[inner sep=0] {\raisebox{0.1cm}{\includegraphics[height=4.4truecm,width=4.0truecm,angle=0,trim={0.0cm 0.0cm 5.5cm 0.0cm}, clip]{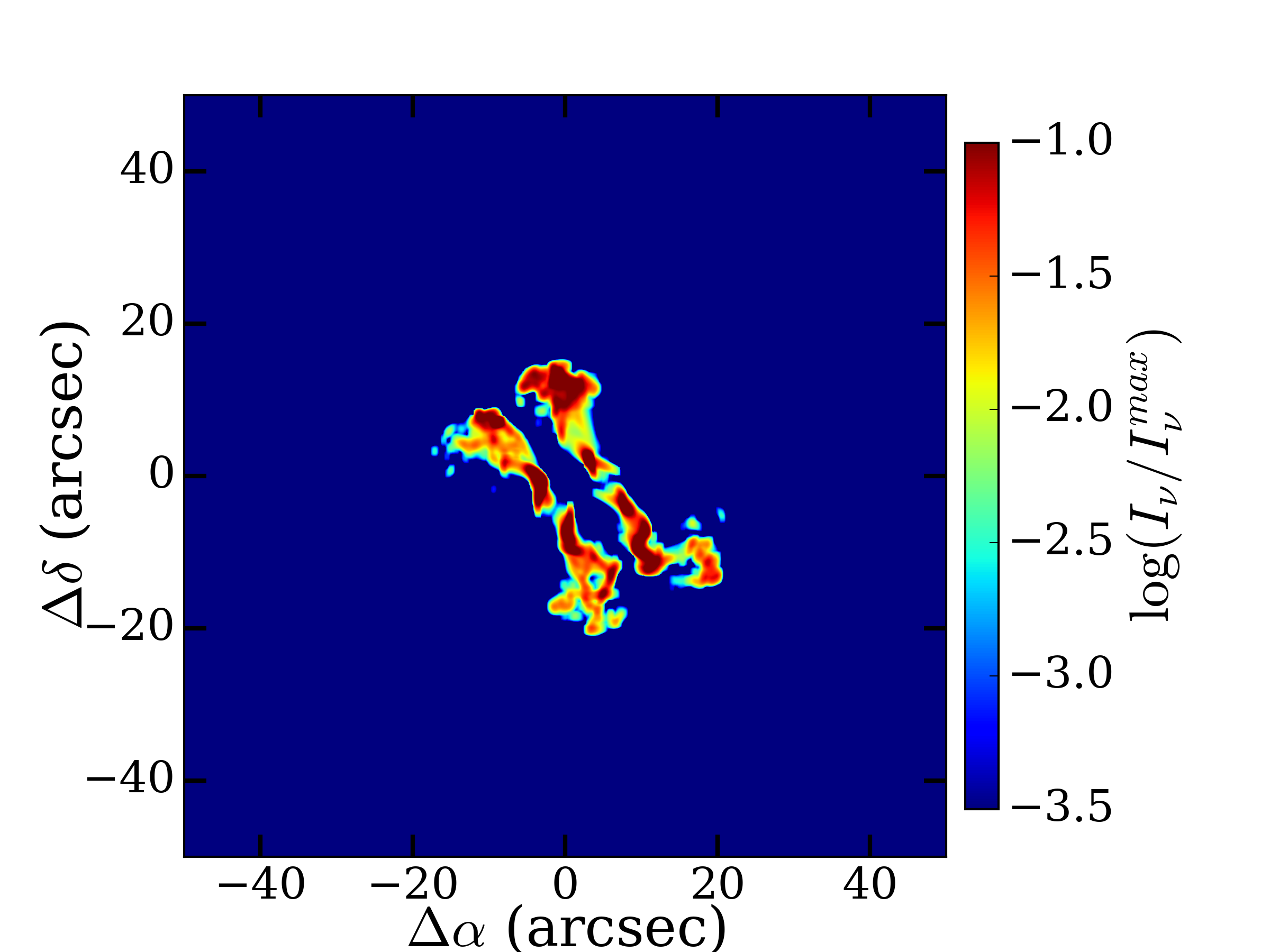}}};
\draw (1.1, 1.5) node[text=white] {\small 73 Myr};
\end{tikzpicture}
\begin{tikzpicture}
\draw (0, 0) node[inner sep=0] {\raisebox{0.1cm}{\includegraphics[height=4.4truecm,width=3.2truecm,angle=0,trim={3.2cm 0.0cm 5.5cm 0.0cm}, clip]{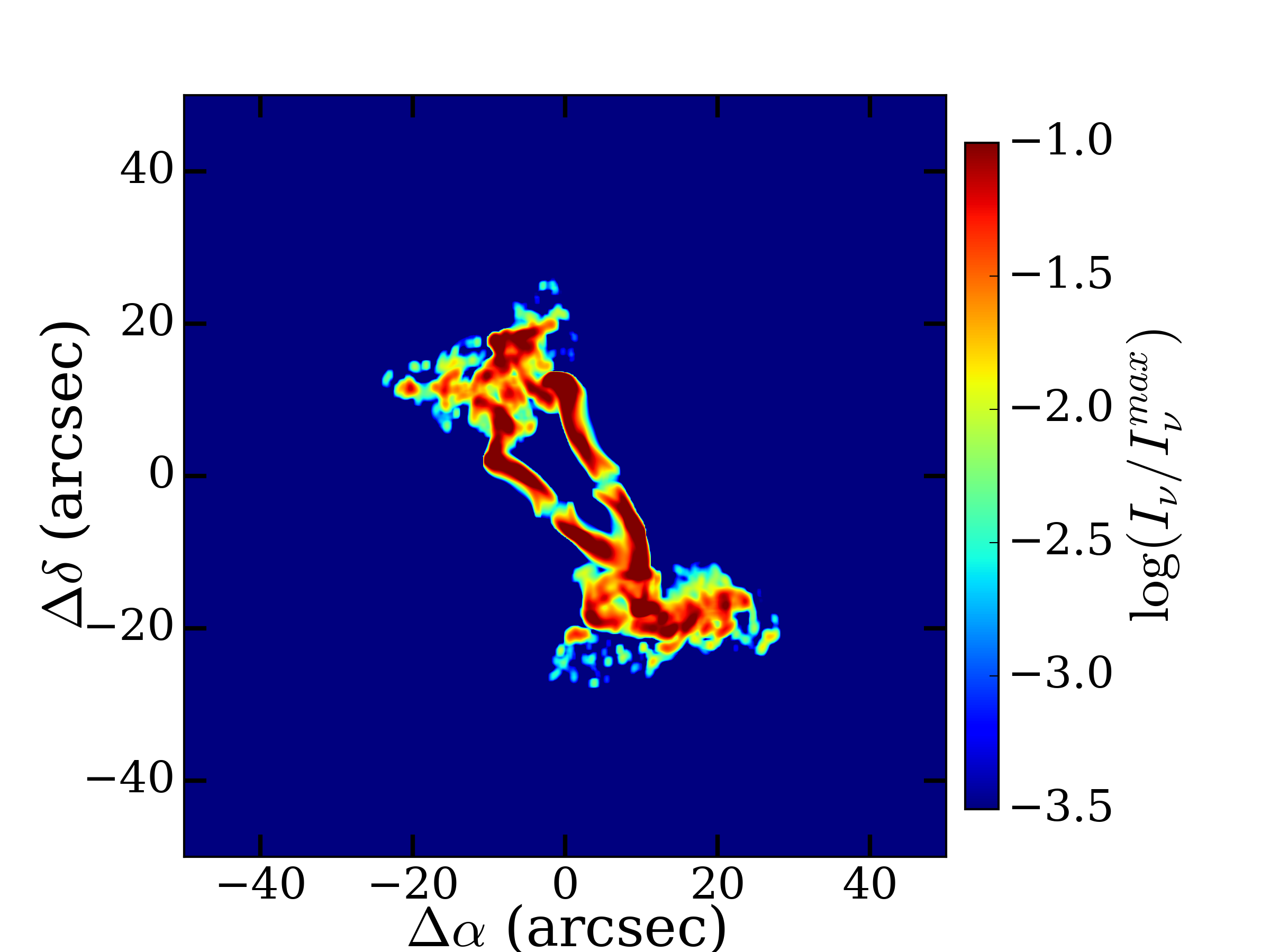}}};
\draw (0.6, 1.5) node[text=white] {\small 150 Myr};
\end{tikzpicture}
\begin{tikzpicture}
\draw (0, 0) node[inner sep=0] {\raisebox{0.1cm}{\includegraphics[height=4.4truecm,width=4.6truecm,angle=0,trim={3.2cm 0.0cm 0.0cm 0.0cm}, clip]{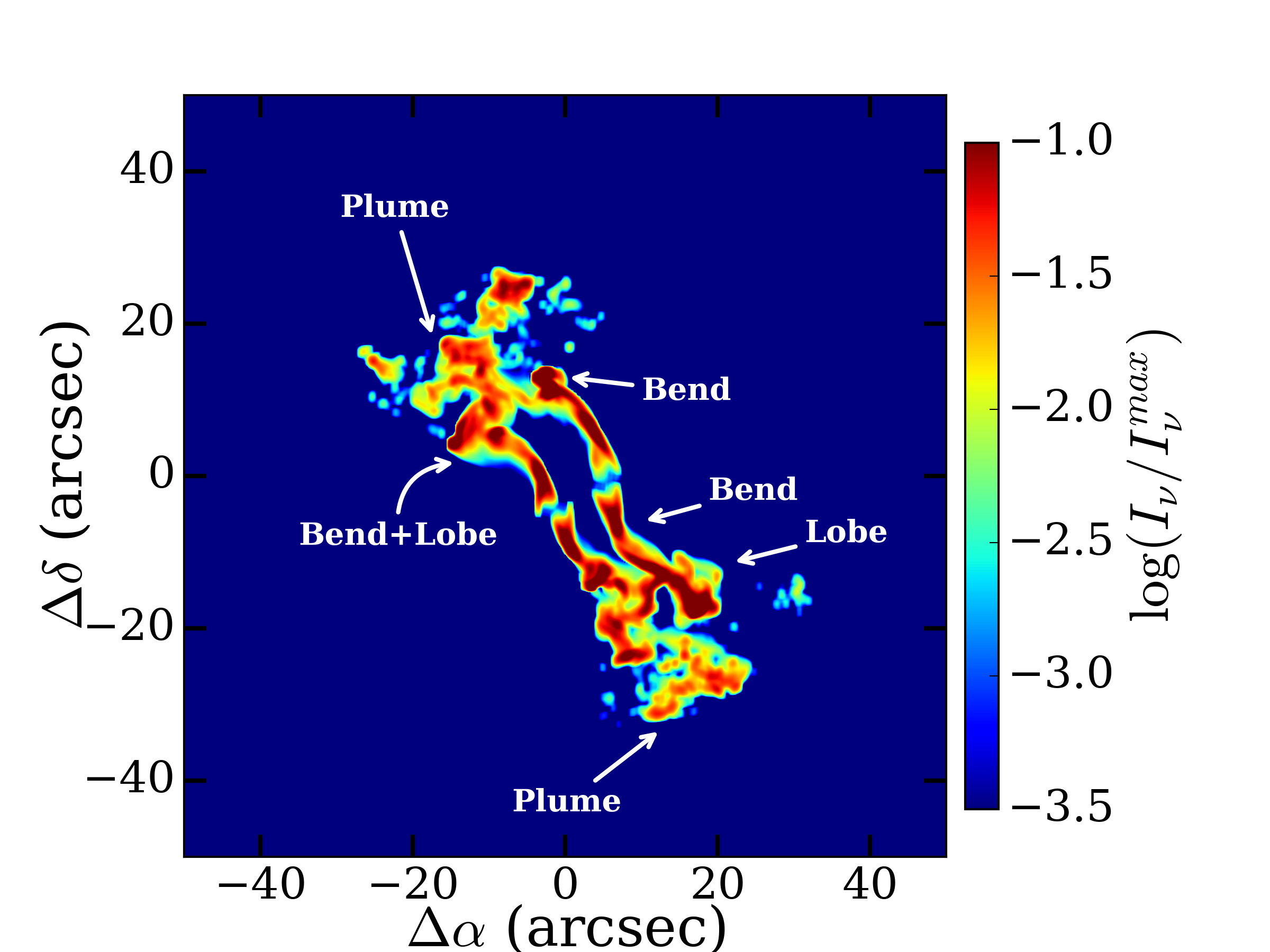}}};
\draw (0.0, 1.5) node[text=white] {\small 190 Myr};
\end{tikzpicture}
\caption{The leftmost panel reproduces the color-coded uGMRT map of the TRG J104454 at 1.4 GHz  \citep{Gopal-KrishnaEtal2022MNRAS.514L..36G}; (the image is used with authors' permission and does not violate copyright). The remaining three panels display the normalized synchrotron intensity maps of the simulated TRG at 1.4 GHz for the model S1, obtained using emissivity data-cubes of both jets.  To produce the mock-intensity maps we use values of $\alpha =0.6$ (see Eqn.\ 4). The approximate times of these simulation snapshots are shown in the respective images. The coordinates of the nuclei of the right and left bipolar jets in the simulation are respectively \{15, 0, 0\} kpc and \{-15, 0, 0\} kpc. The simulated high-resolution maps have been convolved to a lower resolution using a GaussianBlur filter, with the filter size matching the uGMRT beam size. The last panel also marks the structural features of the jets.
}
\label{fig:SimSetup1}
\end{figure*}

\section{Results and Discussion}
The left panel of \autoref{fig:SimSetup1} shows the color-coded uGMRT map of the TRG J104454 at 1.4 GHz \citep{Gopal-KrishnaEtal2022MNRAS.514L..36G}. The other three panels of \autoref{fig:SimSetup1} show the simulated synchrotron intensity maps for the simulation S1 (\autoref{table:RunPars}) using the reference resolution at three different epochs during the jet evolution. 
The synchrotron intensity, $I_\nu$, is approximated using the relation \citep[e.g.,][]{MioduszewskiEtal1997ApJ...476..649M,Antas2024},
\begin{equation}
    I_\nu \propto \delta^{\alpha+2}\left(\frac{P}{\Gamma-1}\right)^{(\alpha+3)/2} \nu^{-\alpha},
\end{equation}
where: $\nu$ is the frequency at which the observation was made; $\alpha$ is the spectral index; $\delta$ is the relativistic Doppler boosting factor, which is $\simeq 1$ for the sub-relativistic jet considered in this work; $P$ is the jet pressure; and $\Gamma$ is the adiabatic gas index. We adopted the synchrotron spectral index as $\alpha = 0.6$, typical for active radio jets \citep[e.g.,][]{JarvisEtal2019MNRAS.485.2710J,SilpaEtal2022MNRAS.513.4208S}. We then integrate the emissivity data-cube along the line-of-sight to obtain the intensity maps. We are viewing the structure along the $z$-axis (a tracer threshold of 0.5 is chosen to consider only emission from the zones predominantly comprised of radio source plasma). All snapshots are rotated counterclockwise by $30^\circ$ with respect to the sky-projected $y$-axis for better visualization and comparison with the observed source. 

Both bipolar jets are found to remain relatively straight with little wiggling during the early stages of their evolution (see the $\sim73$ Myr snapshot in \autoref{fig:SimSetup1}).  However, at later stages, they diverge from symmetrical evolution due to differences in their precession parameters, which influence variations in their longer-term structures. During their growth, the two bipolar jets approach each other and thereby appear to influence their mutual dynamics and show more wiggling of the diffused material, as seen in the snapshot at time $\sim150$ Myr. At later stages, they lose collimation and form diffuse radio lobes, as shown in the right panel of \autoref{fig:SimSetup1}, which also resembles the observed radio map of TRG J104454. The simulation timescale ($\approx 190$ Myr) that matches the observed morphology is in rough agreement with the orbital timescale of the system, which is $t_{orb}=2\pi (d^3/GM_\ast)^{1/2}\sim 230$ Myr, where $d$ is the half-separation between two galaxies, and $M_\ast$ is the total stellar mass $\sim5.4\times10^{11}$ M$_\odot$ \citep{Gopal-KrishnaEtal2022MNRAS.514L..36G}. However, the simulated jet precession timescales required for the observed jet wiggling seen in the uGMRT map are substantially shorter than this orbital timescale, and a possible origin of such timescales is discussed later in Sec.\ 3.3.

\subsection{Jet dynamics}
As soon as the jets start propagating through the ambient medium in the denser core regime, their growth rate and dynamics are affected by the density profile of the ambient medium \citep[e.g.][]{GopalWiita1987MNRAS.226..531G}. We note that the present assumption that the kinetic power of the jet is dominated by the synchrotron plasma in sub-relativistic bulk motion does not exclude the possibility that the primary source of the jet power lies in a relativistic spine.

\begin{figure}
\centering
\includegraphics[height=3.8truecm,width=4.2truecm]{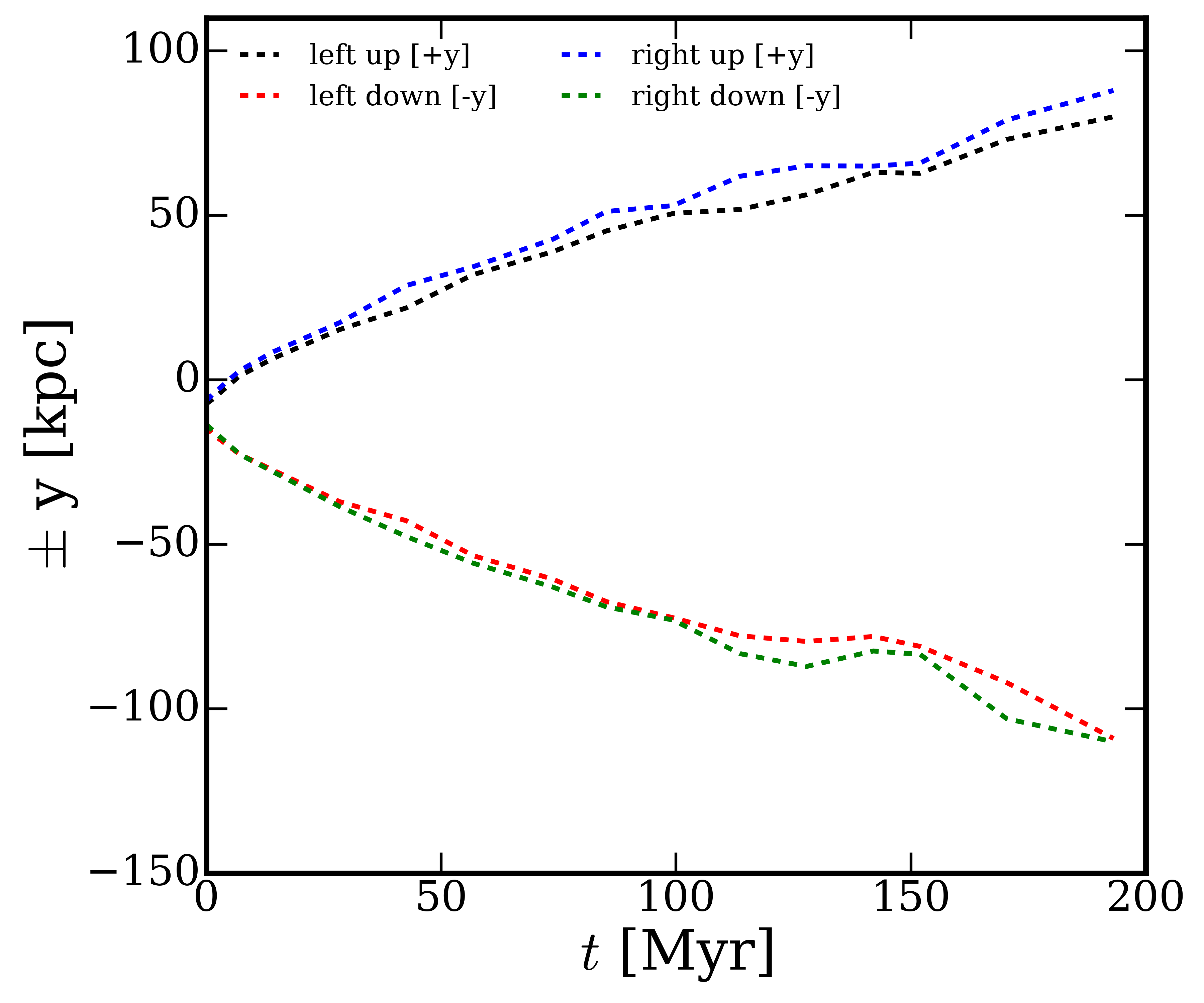};
\hspace{-0.3cm}
\includegraphics[height=3.8truecm,width=4.2truecm]{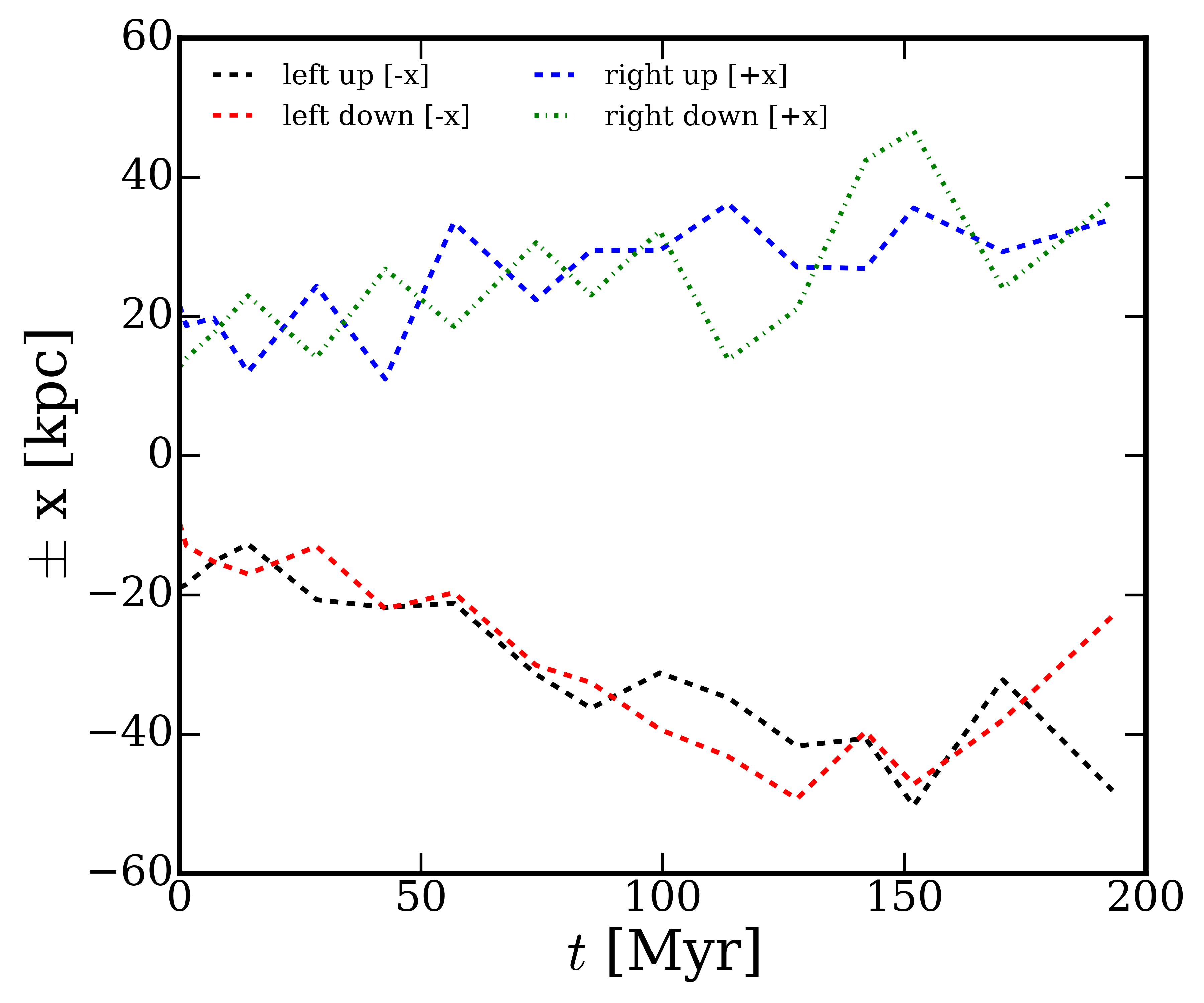};
\caption{Temporal evolution of the jet head for all four jets in the $x-y$ plane, for the setup S1. The left and right panels show approximate $\pm y$ (linear) and $\pm x$ (lateral) positions of the jet heads. Both panels follow the same color conventions to indicate the four jets.
}
\label{fig:JetHead}
\end{figure}

We have measured the jet head evolution for all four jets using $x-y$ slices at different times. A tracer threshold of $10^{-4}$ is used to delineate the full dynamical extent of the lobes. The left panel of \autoref{fig:JetHead} shows the head positions (along the $\pm y$ axes) for all four jets (left jet's upward arm in black, left jet's downward arm in red, right jet's upward arm in blue, and right jet's downward arm in green) during their temporal evolution, which was generated by the propagation of the jets through the denser ambient medium. The expansion of the upper jets (toward the $+y$ axis) is slightly slower than that of the lower jets, which might explain the wider lobe formation by the jets along this direction compared to the $-y$ axis. The northern jets ($+y$-axis) and southern jets ($-y$-axis) exhibit somewhat different evolutionary patterns which could be due to the mutual dynamical effects of the tilted bipolar jets. This results in varied impacts across different sections, further emphasizing the independent evolution of the jets. The right panel of \autoref{fig:JetHead} shows the lateral movement of the jet heads. The same color convention as in the left panel has been followed. It can be seen that, at the initial breakout stage, all four jets had small lateral spreads, which, however, increased at later times.  This can be understood because of their higher thrust and more precise directionality at the time of launch compared to later times and the effect of precession. At the initial phase of their launch, the heads of all four jets are spatially distinct but as time progresses the lobes begin to inflate and make contact among themselves, enhancing a greater lateral spread of the diffused material.   

\begin{figure}
\centering
\includegraphics[width=0.8\columnwidth]{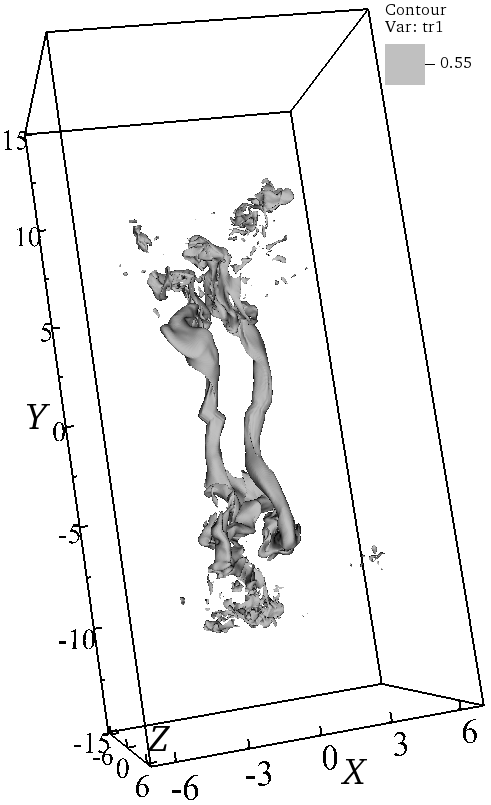};
\caption{3D gray-scale map of the tracer value (tracer values $> 0.5$) illustrating the primary jet structure, as well as its transition and decollimation into broader plumes. All axes are in the code unit.} 
\label{fig:3D_dynamics}
\end{figure}

Helical structures arising from the precession of both jet pairs diminish at late time due to their apparent mutual dynamical interactions as well as their interactions with the heavier ambient medium. Nonetheless, the helical jet trajectories are still noticeable up to a certain spatial extent. At much later stages of evolution, however, the jets lose their collimated 
structures as they slow down, and start dispersing their material over a wider lobe region. 

Note that although the present simulations cover a range of input parameters, as listed in  \autoref{table:RunPars}, only the setup S1 most closely reproduces the observed TRG morphology. We also performed a similar simulation without incorporating the effect of jet precession; then all four jets were found to drill through the ambient medium keeping their initial injected directions, thereby failing to reproduce the observed jet-lobe structures. Therefore, we are confident that the observed morphology of TRG J104454 requires significant precession to be present in both jets.

To illustrate the evolved dynamical structure in 3D, we have plotted tracer values in 3D (\autoref{fig:3D_dynamics}) representing the dominant jet-lobe structure (tracer values exceeding 50\%) for the $\sim190$ Myr snapshot in \autoref{fig:SimSetup1}. The collimated jets and their helical distortions are clearly visible in this visualization, along with their decollimation into tendril-like, wobbly structures after a certain spatial extent. The evolution of the northern and southern arms of the bipolar flows differ considerably: the northern arms interact and collide with each other, while the southern arms remain distinct, as revealed through a rotating view of the rendered image structure.

It is important to point out that the present simulations do not include magnetic fields. This simplification is partly justified by the fact that the magnetic field energies in such low-power systems are typically much lower than the kinetic energy of the jet \citep{Giri2024}, rendering them dynamically less significant in this context. However, magnetic fields are likely to play a critical role in stabilizing the jets against decollimation and shielding them from ambient matter entrainment, 
potentially enabling the twin jets to persist over greater extents. Nonetheless, the role of magnetic fields is highly dependent on their strength and configuration, as they could also induce instabilities in the jets. This remains an open question and warrants further investigation in future studies.

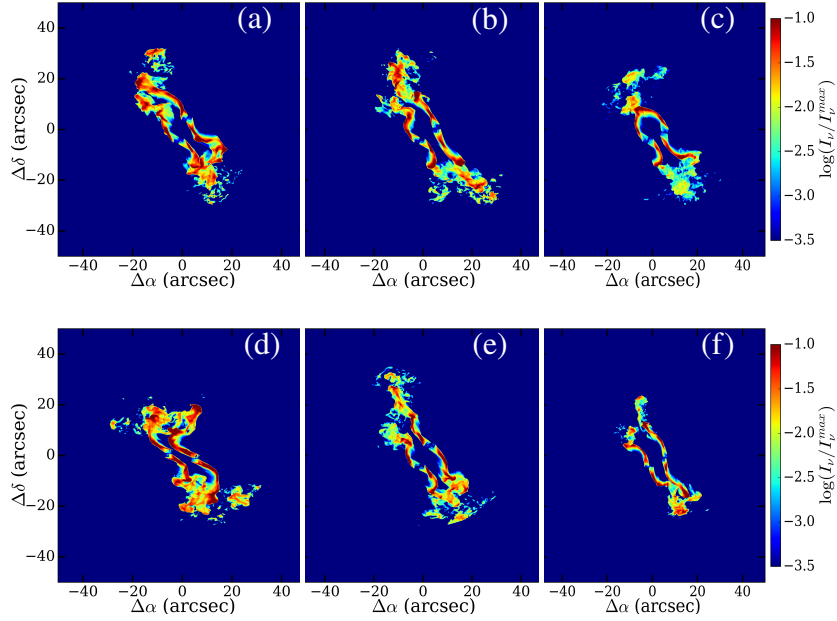
\begin{figure*}[t]
\centering
\begin{tikzpicture}
\draw (0, 0) node[inner sep=0] {\raisebox{0.1cm}{\includegraphics[height=4.2truecm,width=4.0truecm,angle=0,trim={0.0cm 0.0cm 5.5cm 0.0cm}, clip]{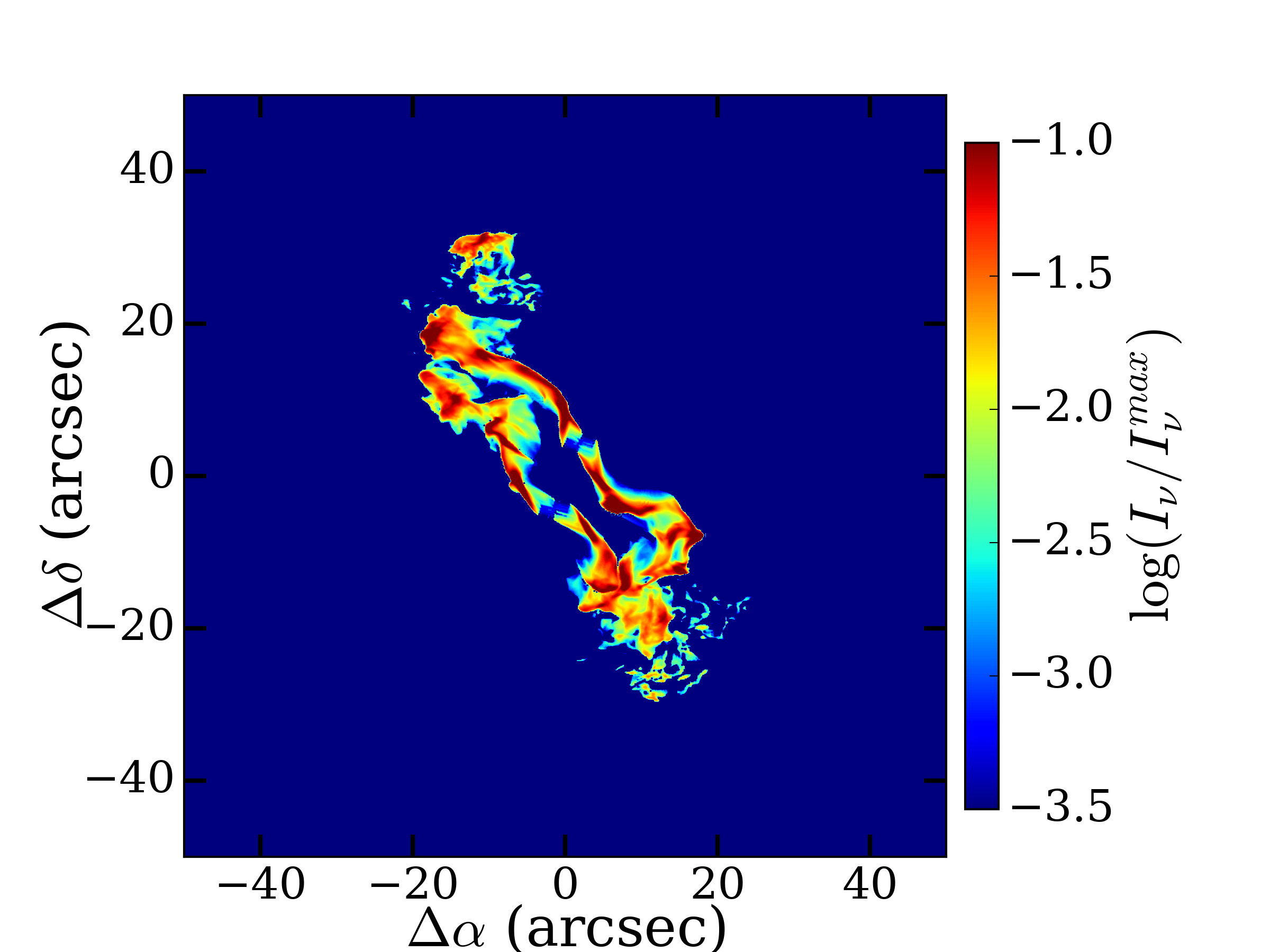}}};
\draw (1.4, 1.5) node[text=white] {\large (a)};
\end{tikzpicture}
\begin{tikzpicture}
\draw (0, 0) node[inner sep=0] {\raisebox{0.1cm}{\includegraphics[height=4.2truecm,width=3.1truecm,angle=0,trim={3.2cm 0.0cm 5.5cm 0.0cm}, clip]{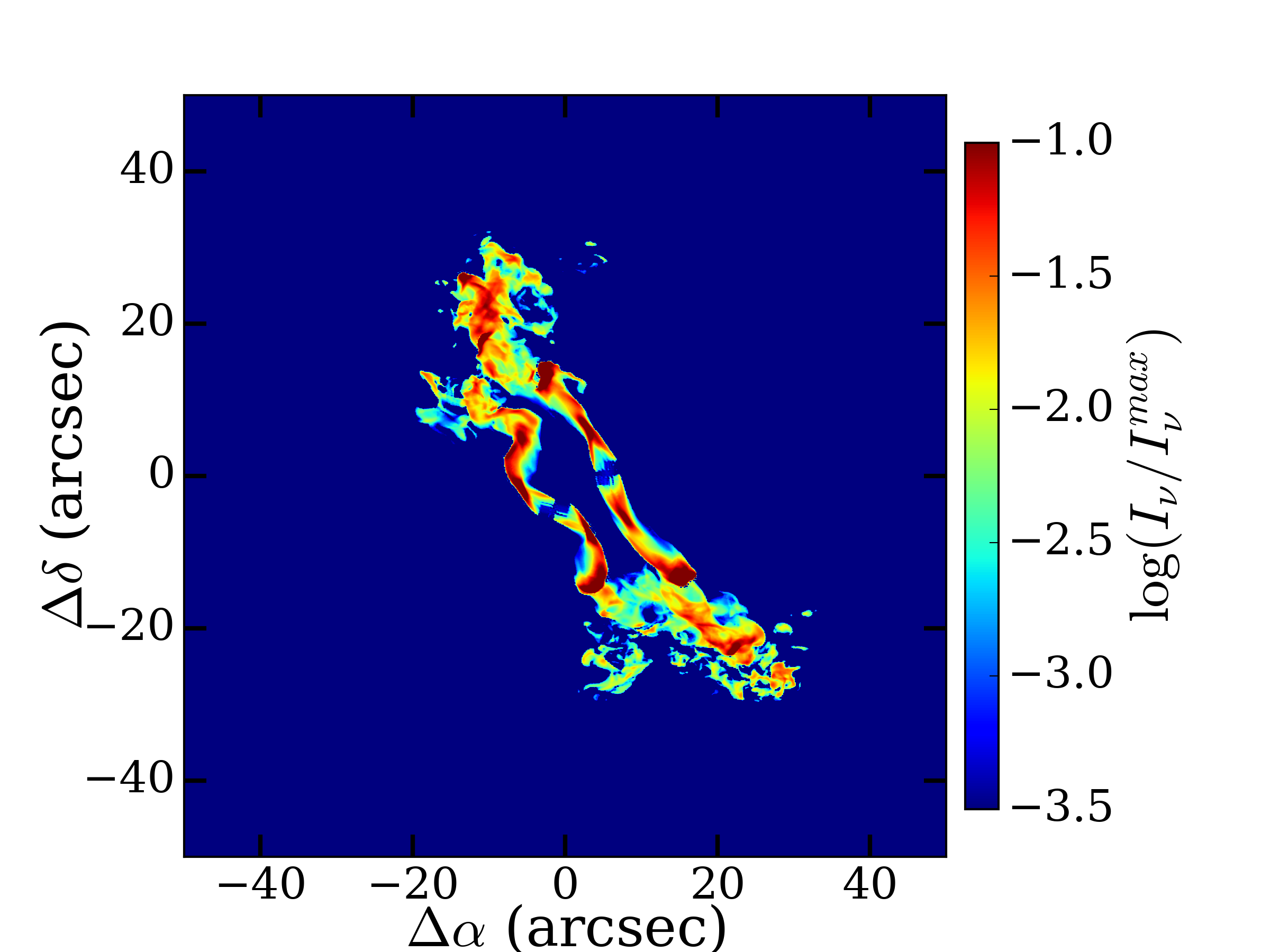}}};
\draw (0.9, 1.5) node[text=white] {\large (b)};
\end{tikzpicture}
\begin{tikzpicture}
\draw (0, 0) node[inner sep=0] {\raisebox{0.1cm}{\includegraphics[height=4.2truecm,width=4.2truecm,angle=0,trim={3.2cm 0.0cm 0.0cm 0.0cm}, clip]{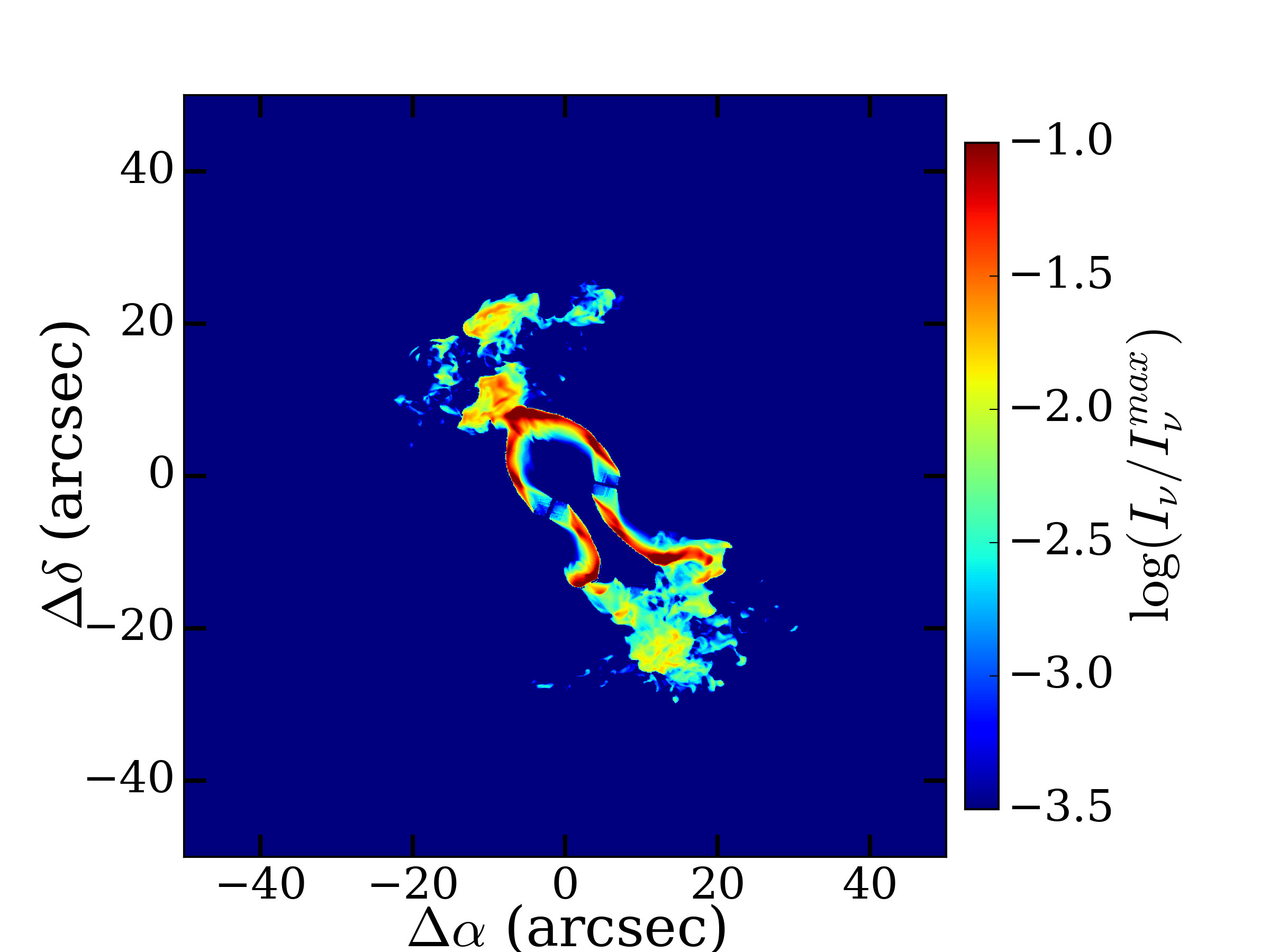}}};
\draw (0.2, 1.5) node[text=white] {\large (c)};
\end{tikzpicture}
\\
\centering
\begin{tikzpicture}
\draw (0, 0) node[inner sep=0] {\raisebox{0.1cm}{\includegraphics[height=4.2truecm,width=4.0truecm,angle=0,trim={0.0cm 0.0cm 5.5cm 0.0cm}, clip]{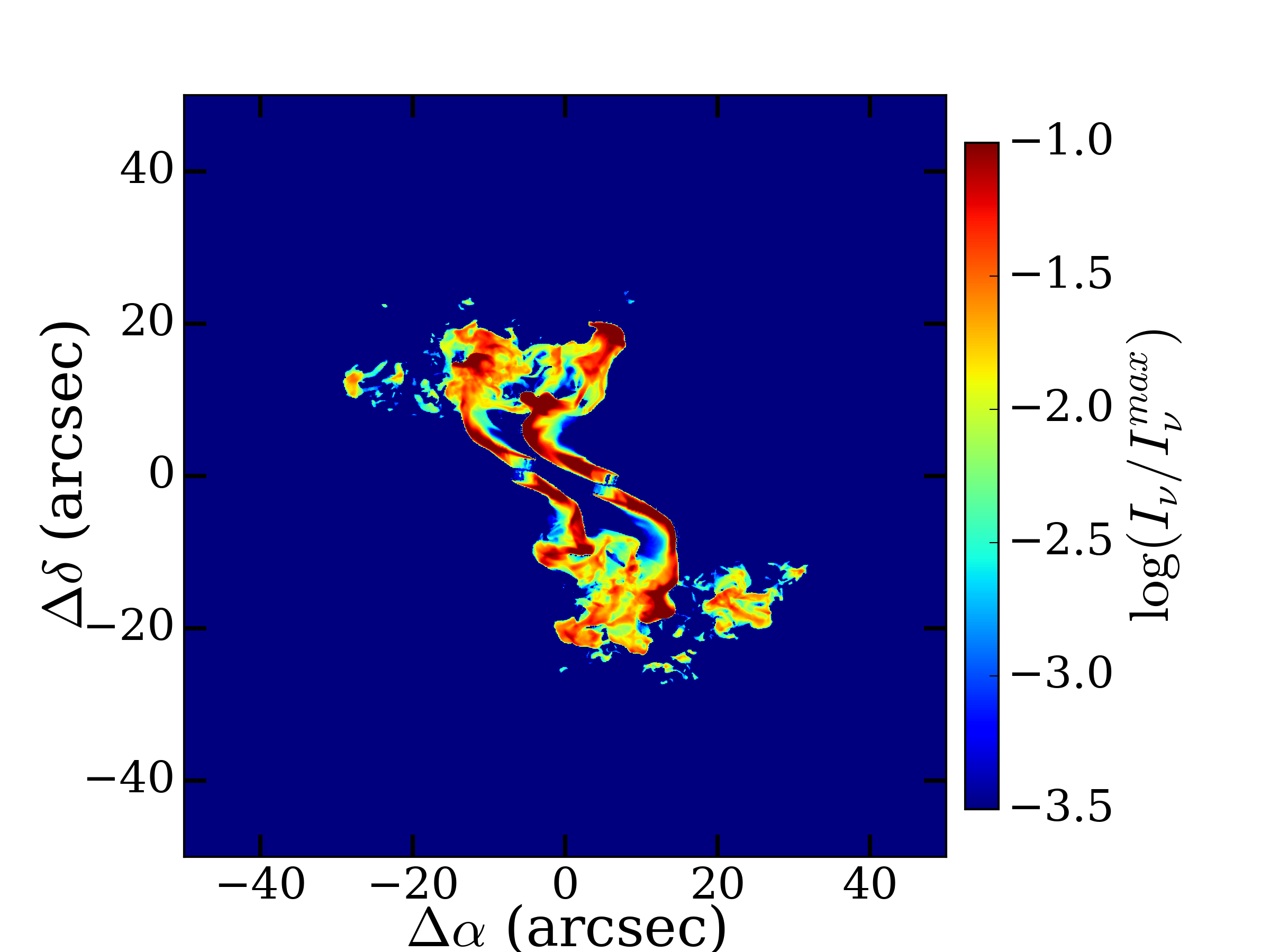}}};
\draw (1.5, 1.5) node[text=white] {\large (d)};
\end{tikzpicture}
\begin{tikzpicture}
\draw (0, 0) node[inner sep=0] {\raisebox{0.1cm}{\includegraphics[height=4.2truecm,width=3.1truecm,angle=0,trim={3.2cm 0.0cm 5.5cm 0.0cm}, clip]{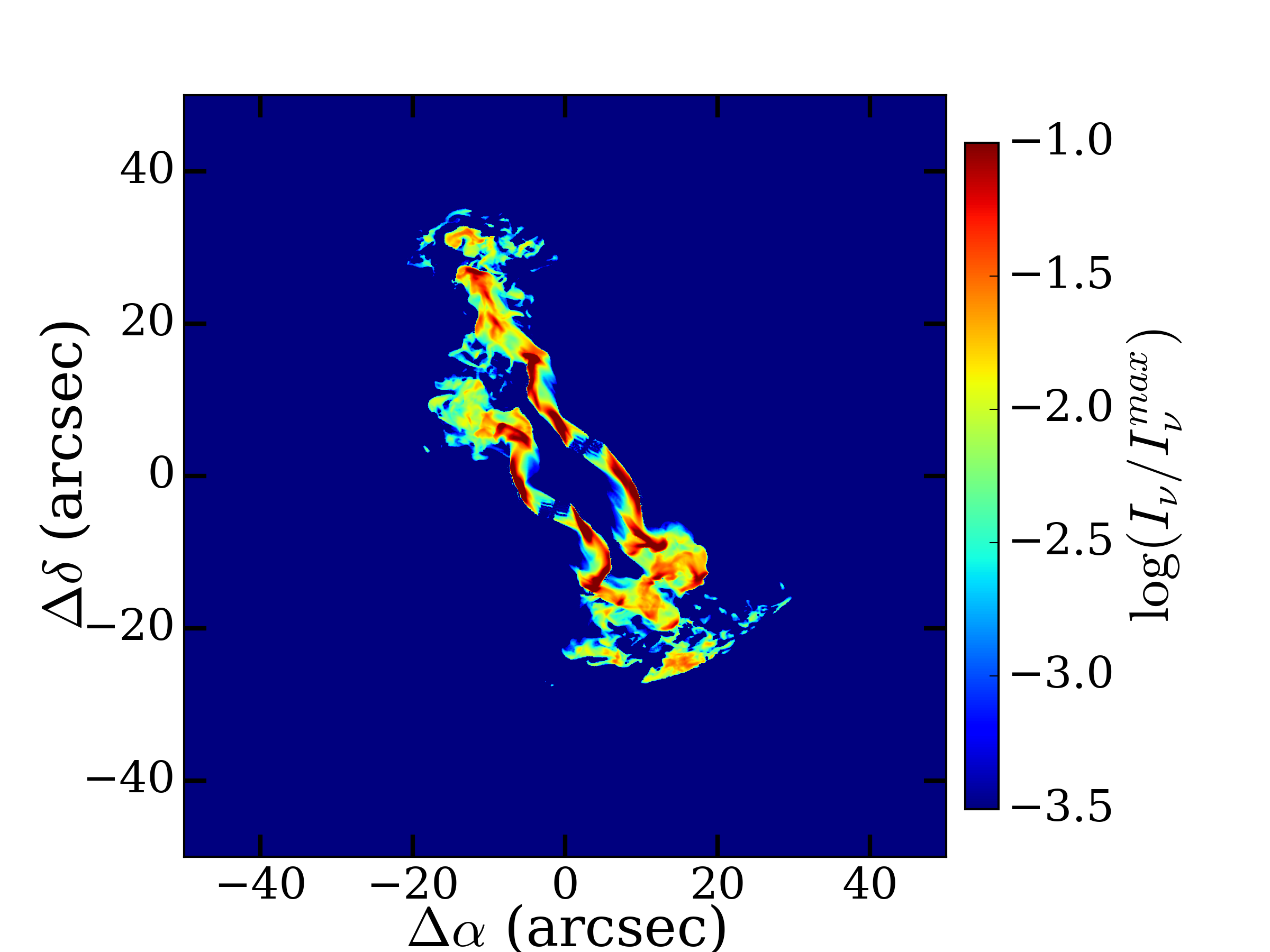}}};
\draw (0.9, 1.5) node[text=white] {\large (e)};
\end{tikzpicture}
\begin{tikzpicture}
\draw (0, 0) node[inner sep=0] {\raisebox{0.1cm}{\includegraphics[height=4.2truecm,width=4.2truecm,angle=0,trim={3.2cm 0.0cm 0.0cm 0.0cm}, clip]{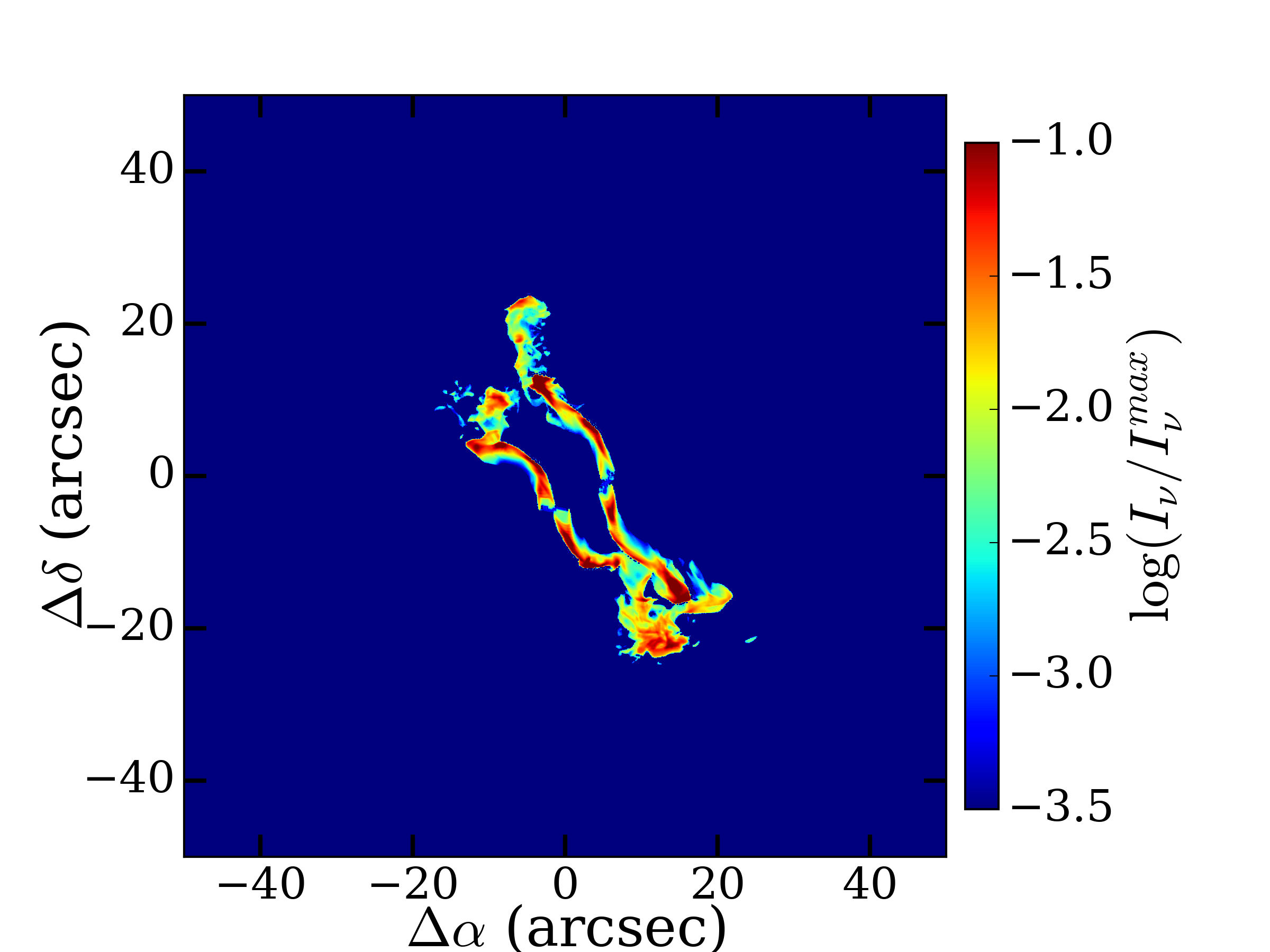}}};
\draw (0.2, 1.5) node[text=white] {\large (f)};
\end{tikzpicture}
\caption{Same as \autoref{fig:SimSetup1}, but for the setups S2--S7 (from (a) to (f)) where the model parameters are varied individually and jointly. Snapshots are taken at 168 Myr (a, b, and e), 189 Myr (c), 176 Myr (d), and 136 Myr (f),  respectively. The adopted jet parameters are given in \autoref{table:RunPars}.}
\label{fig:Setup2to6}
\end{figure*}

\subsection{Jet power and BH accretion mode}
The launching and powering of jets from SMBHs can be related to the properties of the central engine and its accretion states. Observations of jets from black hole X-ray binaries as well as in some AGN provide evidence that jets can be produced in the advection-dominated accretion state and the collapse of the radio jets is observed when the flow transits from the hot to cool state \citep[e.g.][]{FenderEtal2004MNRAS.355.1105F}. However, the collimation and powering of large-scale jets in powerful AGN seem to require the extraction of rotational energy from the BH \citep[BZ process;][]{BZ1977MNRAS.179..433B} and/or the accretion disk through magnetic fields \citep[BP process;][]{BP1982MNRAS.199..883B}. Since these processes can generate a wide range of jet power,  different types of radio sources can be understood within this general framework.  

The SMBHs ejecting radio jets in our simulations have masses of around $1.6\times10^{10}$ M$_\odot$ (left source) and $2.0\times10^9$ M$_\odot$ (right source) \citep{Gopal-KrishnaEtal2022MNRAS.514L..36G}. The $L_{\rm kin}$ values indicated by the simulations suggest that both the SMBHs are accreting in the sub-Eddington regime, with mass accretion rates $\sim 5\times10^{-4}$ and $\sim 4.2\times10^{-3}$ $\dot m_{\rm Edd}$ for the left and right sources, respectively, where the Eddington luminosity, $L_{\rm Edd}=1.3\times10^{38}$ (M$_{\rm BH}$/M$_\odot$) erg s$^{-1}$. For the present sources, where very sub-Eddington accretion is indicated, advection-dominated states are indeed expected \citep{NarayanYi1994ApJ...428L..13N}. 

\subsection{Origin and timescale of jet precession}
Precession of radio jets (either bipolar or twin bipolar) may occur due to several underlying processes including magnetic torques, warped disks, and gravitational torques in a binary \citep{PizzolatoSoker2005AdSpR..36..762P}. In a binary SMBH system, if either of the nuclei is actively accreting and generating radio jets, then the precession is induced around the orbital axis as the galaxies start to merge and the nuclei come closer to form a bound spiraling orbit \citep{BegelmanEtal1980Natur.287..307B}. Such processes can misalign the binary's orbital axis and the SMBH spin axis. \citet{GowerEtAl1982ApJ...262..478G} explored models of precessing twin-jets for several observed systems for a range of bulk jet velocities that demonstrated the effect of orbital motion. Even without precession, spiraling jets can be formed from the combined effects of jet velocity and the orbital velocity of the system. Interestingly, 3C 315 in their sample required a longer precession period ($\sim 10$ Myr) to match the observation, which was quite similar to our simulated precession periods. The radio jets in two galaxies separated by a few tens of kpc can be affected by the dynamical interactions arising from the large-scale galactic gas distribution and their orbital velocities. Occasionally, high-velocity circular orbits can even produce a WAT \citep[e.g., 3C 465,][]{WirthEtAl1982AJ.....87..602W}, as was observed in the TRG 3C 75 \citep{OwenEtal1985ApJ...294L..85O}.

Alternatively, precession can be associated with accretion disk instabilities such as Lense–Thirring (LT) or the related Bardeen–Petterson effects \citep{BardeenPetterson1975ApJ...195L..65B}. The accretion disk may start precessing due to these instabilities if its angular momentum is misaligned with the angular momentum of the central spinning SMBH. Moreover, if the orientation of the disk determines the jet axis, then this would induce a precession in the jet axis \citep[see ][for some general relativistic simulations]{LiskaEtAl2018MNRAS.474L..81L}. 

The precession period of the jets in an isolated AGN or a binary system can cover a broad range from a few years to a few tens of Myr. A wide range of precession timescale results depending on the scale of LT precession of the central SMBH due to the misalignment of the spin axis of the SMBH with the direction of the angular momentum vector of the accreting matter \citep{ScheuerFeiler1996MNRAS.282..291S}. Short precession periods of 11 yr and 7 yr were estimated for the cases of M87 and M81, which could have arisen from the misalignment of the black hole spin axes with the tilted disks \citep{CuiEtAl2023Natur.621..711C,vonFellenbergEtAl2023AA...672L...5V}. However, in the case of large-scale structures, e.g., in galaxy clusters, the precession period can be a few tens of Myr as was inferred for 3C 84 \citep{DunnEtAl2006MNRAS.366..758D,Falceta-GoncalvesEtAl2010ApJ...713L..74F}. These authors assumed that the precession was induced by the LT effect at a larger length scale, which is more likely since the accretion disks are connected with the merger processes and, therefore, likely to be misaligned with the SMBH spin axis.
On the other hand, Hydra A can be modeled with a shorter precession timescale of just $\sim1$ Myr \citep{NawazEtAl2016MNRAS.458..802N}.  

We have estimated the jet precession period in this work assuming that the LT effect allows the disk to warp at larger radii. The precession period can be written as \citep{Pringle1992MNRAS.258..811P,ScheuerFeiler1996MNRAS.282..291S},

$$
    P_j\sim\frac{\nu}{\dot M} \left(\frac{a_k cM_{\rm BH}}{2\nu G}\right)^{\frac{1}{2}},
$$
where, $\nu =\alpha a_s H$, is the disk kinematic viscosity, which is estimated using Shakura-Sunyaev prescription \citep[SS73 disk,][]{ShakuraSunyaev1973A&A....24..337S}, and $G$, $c$, and $\dot M$ are the universal gravitational constant, the speed of light, and the disk mass accretion rate, respectively. The sound speed $(a_s)$ is calculated as $\sqrt{k_BT/m_p}$ for a typical cold disk temperature of $3\times 10^4$ K. The scale height of the SS73 disk $(H)$ is given by $a_s/\varv_\phi$, where $\varv_\phi$ is the angular velocity of the flow. For a standard warp radius, $R_{warp}$ of 150-1000 $r_g (=GM_{\rm BH}/c^2)$ \citep{ScheuerFeiler1996MNRAS.282..291S}, BH spin $a_k=0.8$, $\alpha = 0.2$ \citep[][see for jetted sources]{XieEtAl2009ApJ...707..866X,MondalEtal2022AA...663A.178M}, and $M_{\rm BH}$, and mentioned $\dot M$ in Sec 3.2, the precession timescale is in a range of $\approx7-50$ Myr, which is consistent with the timescales used in our simulations. The estimated timescales also match with the viscous timescale $(t_\nu=t_{dyn} (H/R_{warp})^{-2}/\alpha )$ of the accreting matter, where $t_{dyn}$ is the dynamical timescale of the infalling matter. We thus surmise that LT precession may significantly warp the disk at rather large radii, where the inward advection is balanced by the viscous torque. Therefore, the jet launching and energy extraction to power the jets is more like a BP scenario. This can produce precession in large-scale radio jets with timescales of a few tens of Myr. This supports the shorter precession periods of $\sim 7-50$ Myr for the jets inferred in this work, compared to the full dynamical timescale ($\approx 190$ Myr), which produces agreement with the observed uGMRT morphology of TGR J104454.

In addition to uGMRT morphology, the present system also showed a larger arc-shaped morphology in LoTSS-DR2 data \citep[see][]{Gopal-KrishnaEtal2022MNRAS.514L..36G}, which is mostly diffused emission. It has been discussed by \citet{NoltingEtAl2023ApJ...948...25N} that such large arc morphologies in jets may originate from a longer precession timescale and the drop off in the surface brightness of the synchrotron plasma further away from the original jet direction. Due to this, the precession effects may no longer be detectable. Such a scenario can occur if the AGN activity persists for a time period of $>100$ Myr \citep[see][]{TurnerShabala2015ApJ...806...59T}. We performed a simulation with a much longer precession period, which produced a larger arc in the original jet (see AS15 for these parameters in \autoref{table:AddRuns} in Appendix B). We note that orbital motion alone is unlikely to be able to produce the observed small-scale bends and lobes in the jets.

\subsection{Search through the parameter space}
We have verified that the simulated TRG morphology significantly deviates from the observation once the relevant model parameters depart substantially from the model S1, as illustrated by the results for the models S2--S6, shown in \autoref{fig:Setup2to6}. Particularly large deviations from the observed jet morphology are seen for S4 and S5 shown in \autoref{fig:Setup2to6}(c) and \autoref{fig:Setup2to6}(d) when both bipolar jets are assigned different velocities. The initially higher-velocity jets are found to eventually advance more slowly. Furthermore, these jets also showed larger arcs in their jet morphology. These features of the sub-relativistic jets may be understood as arising primarily from the precession they undergo: faster jets make larger excursions sooner from their initial directions. This would result in their forward thrust spreading over a larger area which can partially disrupt the jets, reshape them, and induce instabilities, thereby accelerating the transfer of the jet's kinetic energy into the ambient medium. Additionally, higher-velocity jets have lower mass loading, so that they more quickly accumulate equal amounts of swept-up mass, which slows them down \citep{MondalEtal2022MNRAS.514.2581M}.

Varying the inclination angle between the two bipolar jets can also affect the jet morphologies, as shown in other panels (a, b, and e) of \autoref{fig:Setup2to6}. We see that for any of the substantial deviations in parameter values from those of S1 shown in \autoref{fig:Setup2to6}(a-e), the simulated TRGs deviate substantially from the observed morphology, although model S5 (panel d) comes fairly close. This makes it clear that numerical simulations serve as an effective tool for constraining the parameter space responsible for generating such complex structures, although they cannot provide unique solutions. Simulations of jets with narrower radii (S7) have also been tested. While there is a coarse agreement with the observed morphology, this simulation neither produces the wider lobe in the top left jet nor the details of the wiggling pattern observed in this jet system (see \autoref{fig:Setup2to6}f). Such a consistency check further implies that the observed morphology depends not only on the jet-jet interaction but also on the internal dynamical parameters of the individual jets.
Given the multi-parameter nature of the system, we considered the possibility that different combinations of precession parameters might also reproduce a similar TRG morphology. To test this, we conducted an additional series of 15 simulations (AS1–AS15), exploring a broader parameter space (see \autoref{table:AddRuns} in Appendix B). However, reassuringly, none of these simulations yielded a better match than S1.

\section{Conclusions}
In this work, we have performed non-relativistic hydrodynamic 3D simulations of two bipolar jets emanating from a gravitationally bound pair of host galaxies and propagating through a declining density external medium. Our simulations were aimed to reproduce the observed morphology of the twin radio galaxy TRG J104454+354055. The main results obtained are:

\begin{itemize}
    \item Injecting both bipolar jets without precession results in the jets simply drilling through the medium, without forming the observed helical or wiggling structures. This inconsistency with the observed radio morphology, underscores the critical role of jet precession in explaining the morphology of TRG J104454. This further demonstrates that the precession can play a crucial role in the dynamical evolution of the jet pairs, even when their central engines are separated by (a few tens of) kpc distances.

    \item Low-velocity jets with different inclination angles and precessional periods can satisfactorily reproduce the observed morphology of the TRGJ104454.

    \item For the optimal choice of input parameters among those we considered, the jet propagation along the $\pm y$-axis was steady, while much random motion was observed in the $\pm x$ directions in the $x-y$ plane. This arises from the effects of precession and dynamics of the two bipolar jets.

    \item Our study suggests the AGN are in an advection-dominated accretion mode with a preference for the BP mechanism for the jet launching. 
    
    \item The simulated jet precession timescales are substantially shorter than the full dynamical timescale of the jets and the orbital timescale of the system. Such precession timescales could arise from the LT effects due to misalignment of the black hole's spin axis from that of the accretion disk feeding it.
\end{itemize}
    
 In summary, the preferred simulation reproduces the observed complex morphologies of twin radio jets in TRG J104454 and underscores the significant role of precession in low-velocity jets. Moreover, these simulations highlight the importance of jet-ambient interactions in shaping their morphologies on a large scale in relatively undisturbed environments. Similar simulations can be used to probe the nature of any new Twin Radio Galaxies with similar morphologies expected to be revealed by the next-generation telescopes, such as SKA and its pathfinders. 
 
 However, not all TRGs necessarily exhibit precession; as seen in other discovered TRGs, external influences such as cluster winds likely play a key role. For instance, 3C 75 has been successfully modeled under the influence of cluster winds \citep{MusokeEtal2020MNRAS.494.5207M}, while the morphology of PKS 2149-158 can be attributed to buoyant motion \citep[e.g.,][for a relevant buoyant evolution of a radio jet]{BurnsBalonek1982ApJ...263..546B}. 
 
 Since the present model allows for precession periods and angles as independent parameters for each bipolar jet, it can produce a broad range of jet morphologies. If precession is nonetheless found to be unimportant for future TRGs, the helical nature in their jet morphologies might be explained by the large-scale gas motion in the circum-galactic media associated with the interacting galaxy pair, or the orbital motion of the binary system, or continuous tidal interaction \citep{WirthEtAl1982AJ.....87..602W}, as appear to be required for the two previously known TRGs.

\section{Acknowledgements}
We thank the anonymous referee for making insightful comments and helpful suggestions. SM acknowledges the Ramanujan Fellowship grant (RJF/2020/000113) by DST-ANRF, Govt.\ of India for this research. GG is a postdoctoral fellow under the sponsorship of the South African Radio Astronomy Observatory (SARAO). Financial assistance of the SARAO towards this research is hereby acknowledged (\url{https://www.sarao.ac.za/}). SM and RJ 
acknowledge the use of the computing resources made available by the Computer Centre (NOVA Cluster) of the
Indian Institute of Astrophysics for this work. GK thanks the Indian National Science Academy for a Senior Scientist position during which part of this work was carried out. 
LCH was supported by the China Manned Space Program (CMS-CSST-2025-A09), the National Science Foundation of China (12233001), and the National Key R\&D Program of China (2022YFF0503401).

\section*{Data Availability}
Simulated data can be available upon reasonable request.

\appendix 
\restartappendixnumbering
\section{Resolution Test }\label{sec:appendA}
\begin{figure*}
\centering
\begin{tikzpicture}
\draw (0, 0) node[inner sep=0] {\raisebox{0.1cm}{\includegraphics[height=4.6truecm,width=4.2truecm,angle=0,trim={0.0cm 0.0cm 5.5cm 0.0cm}, clip]{Figures/S6RBent30si23P2Rjp4vp15NoFlip.0012.png}}};
\draw (1.2, 1.5) node[text=white] {\small 168 Myr};
\end{tikzpicture}
\begin{tikzpicture}
\draw (0, 0) node[inner sep=0] {\raisebox{0.1cm}{\includegraphics[height=4.6truecm,width=4.6truecm,angle=0,trim={3.2cm 0.0cm 0.0cm 0.0cm}, clip]{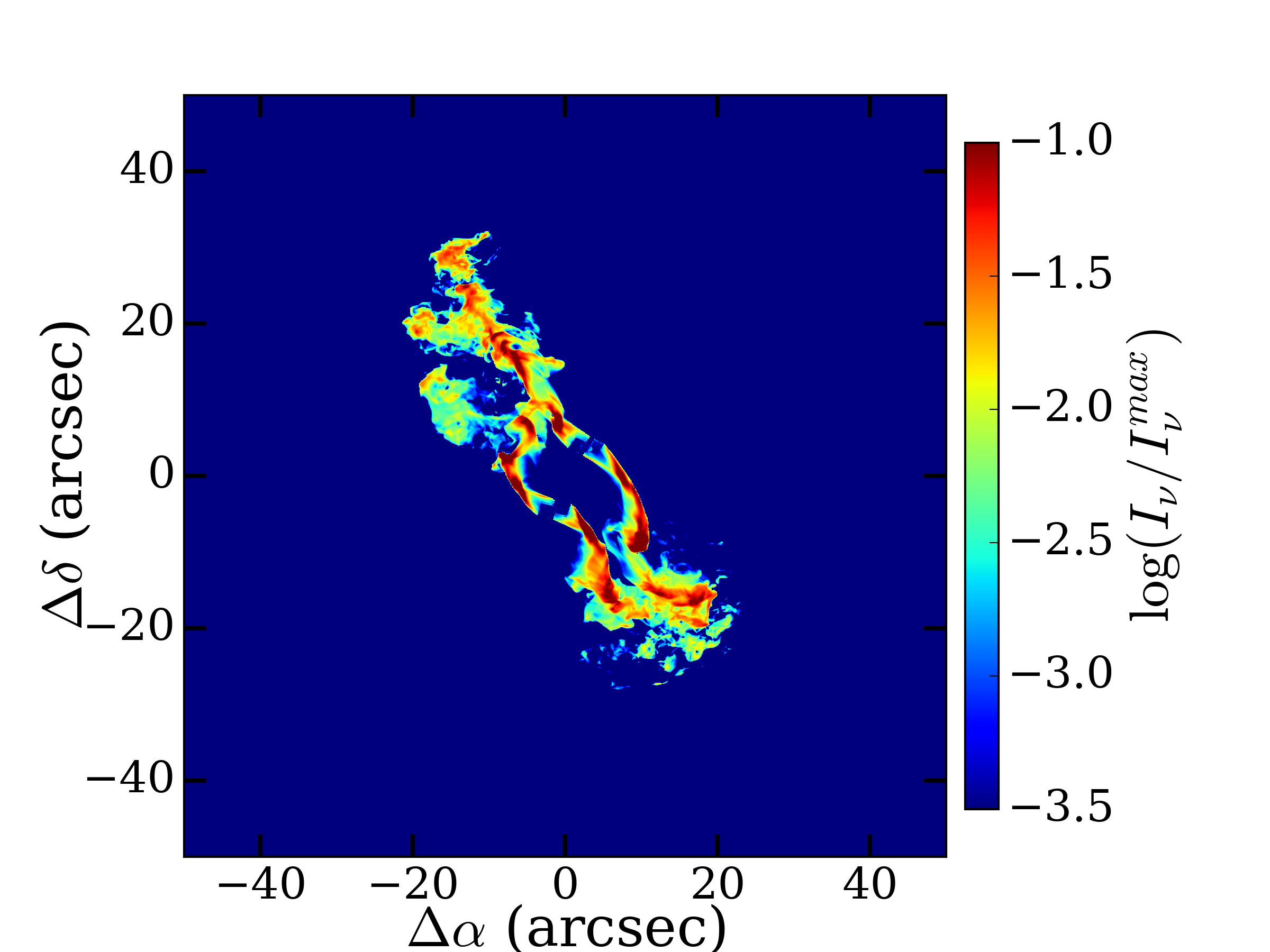}}};
\draw (0.0, 1.5) node[text=white] {\small 168 Myr};
\end{tikzpicture}
\hspace{0.3cm}
\includegraphics[height=4.3truecm,width=5.0truecm]{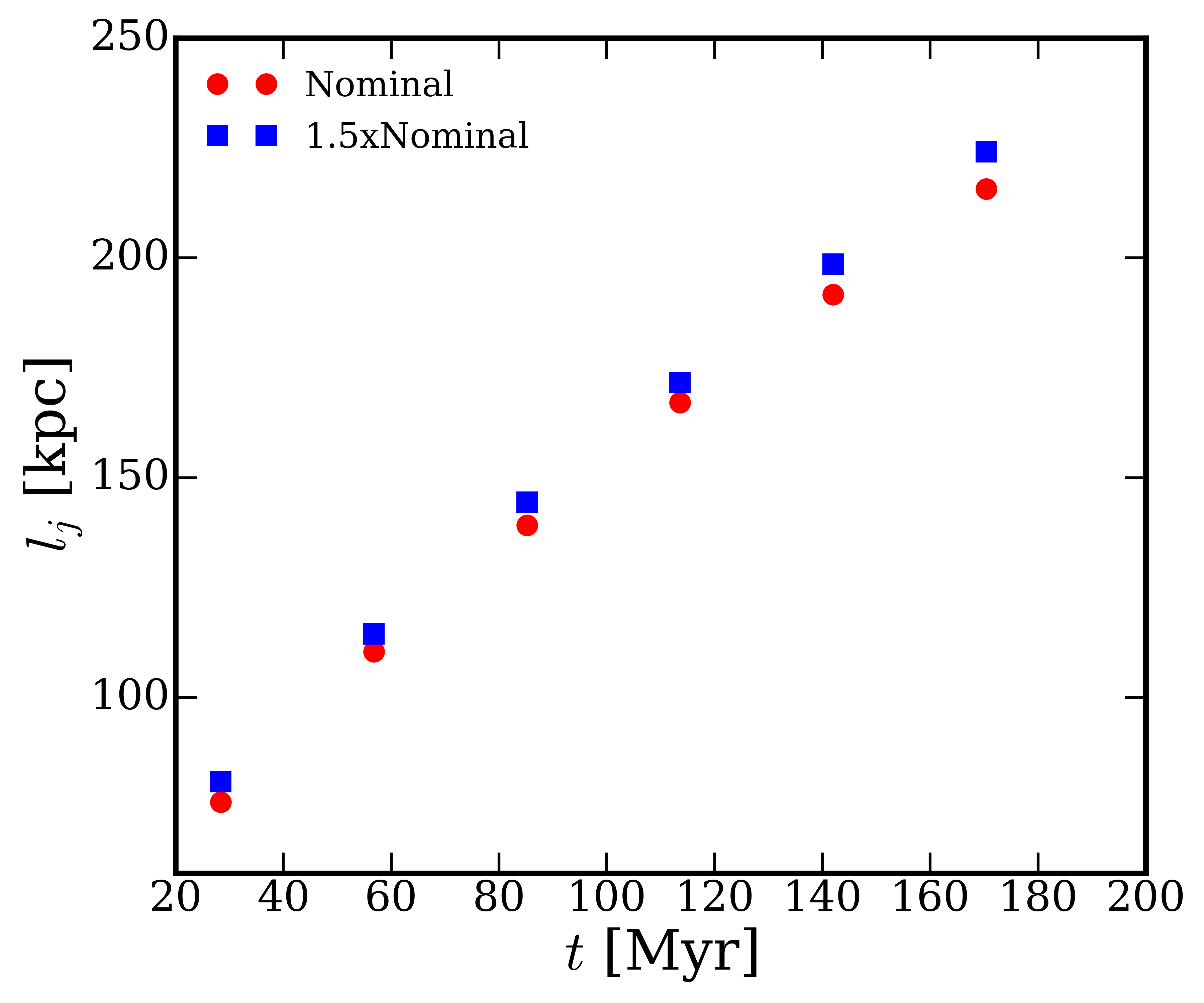};
\caption{Comparison of the emissivity maps for two different resolutions for the setup S6. The left and middle panels respectively show the nominal \{$336\times720\times336$\} and high-resolution \{$504\times1080\times504$\} maps. A comparison of the total length of the jet ($l_j$) for two resolutions is shown in the right panel. The red circles and blue squares correspond to the nominal and high-resolution cases, respectively.
}
\label{fig:ConvTest}
\end{figure*}

The resolution test is performed for the bipolar precessing jets (setup S6) with \{x, y, z\} grids of dimensions \{$336\times720\times336$\} and \{$504\times1080\times504$\}. The left and middle panels of \autoref{fig:ConvTest} show the emissivity maps for both reference and $1.5\times$ reference resolution grids. The right panel shows the temporal evolution of the total jet length ($l_j$), which is estimated for the $\pm y$ length where the first non-zero value of the tracer is obtained. The red and blue points in \autoref{fig:ConvTest} correspond to the reference and $1.5\times$ reference resolutions. As one can see, both resolutions give essentially similar results with  $l_j$ values differing by only $\sim 4\%$. We note that these jet simulations do show different small-scale structures throughout their propagation in the ambient medium, so that aspect may not be very well converged.  However, the similarity of overall extents and morphologies appear to be sufficient for the purpose of this work \citep[see discussions in ][]{DubeyEtal2023ApJ...952....1D}.

\section{Additional simulations}
We ran 15 additional simulations (not shown here) for different combinations of model parameters; however, none of these provided a better match to the observations. The parameters for those simulations are given in \autoref{table:AddRuns}. From those simulations we noticed that a significant difference in jet velocities can almost stall the propagation of the low-velocity jets, while the high-velocity one propagates much faster (e.g., model AS2). Moreover, jets that precess with shorter periods show more frequent curves in their helical structure, end up with diffuse emissions after traveling a shorter distance, and those diffuse emissions start interacting (e.g., in AS5). If two jets have opposite inclinations (as in AS8 and AS12) with positive inclinations of the left jet and a lower velocity compared to the right jet, then the low-velocity left jet remains nearly straight, without wiggling or developing lobes.

\begin{table}
\caption{\label{table:AddRuns} Additional setups ran to search for possible TRG matches with observation.}
\begin{tabular}{ccccccccc}
\hline
Additional & $\theta_l$ & $\theta_r$ &$P_{r}$ &$P_{l}$ &$\Psi_{j}$ &$M_j$&$M_j$&$R_j$\\
setups&($^\circ$)&($^\circ$)&(Myr)&(Myr)&($^\circ$)&(right)&(left)&(kpc)\\
\hline
AS1&-45 &-30 &28 &28 &20 & 65.2&65.2&4\\
AS2&-45 &-30 &28 &28 &23 & 87.0&43.5&4\\
AS3&-45 &-35 &28 &28 &20 & 65.2&65.2&4\\
AS4&-45 &-35 &53 &28 &20 & 65.2&65.2&4\\
AS5&-45&-30&7&7&20&73.9&73.9&4\\
AS6&-45&30&14&14&20&65.2&65.2&4\\
AS7&-45&40&49&28&20&87.0&87.0&4\\
AS8&15&-35&14&14&30&82.6&52.2&4\\
AS9&-30&45&49&49&30&82.6&65.2&4\\
AS10&-40&45&35&35&30&82.6&65.2&4\\
AS11&-40&45&28&28&30&65.2&65.2&4\\
AS12&20&-35&17&17&30&82.6&60.9&4\\
AS13&40&45&49&49&30&87.0&65.2&4\\
AS14&35&35&28&28&30&78.3&65.2&4\\
AS15&-45&40&140&84&20&65.2&65.2&4\\
\hline
\end{tabular} 
\end{table}


\bibliography{reference} 

\begin{thebibliography}{}
\expandafter\ifx\csname natexlab\endcsname\relax\def\natexlab#1{#1}\fi
\providecommand{\url}[1]{\href{#1}{#1}}
\providecommand{\dodoi}[1]{doi:~\href{http://doi.org/#1}{\nolinkurl{#1}}}
\providecommand{\doeprint}[1]{\href{http://ascl.net/#1}{\nolinkurl{http://ascl.net/#1}}}
\providecommand{\doarXiv}[1]{\href{https://arxiv.org/abs/#1}{\nolinkurl{https://arxiv.org/abs/#1}}}

\bibitem[{K.~N. {Abazajian} {et~al.}(2009){Abazajian}, {Adelman-McCarthy}, {Ag{\"u}eros}, {Allam}, {Allende Prieto}, {An}, {Anderson}, {Anderson}, {Annis}, {Bahcall}, {Bailer-Jones}, {Barentine}, {Bassett}, {Becker}, {Beers}, {Bell}, {Belokurov}, {Berlind}, {Berman}, {Bernardi}, {Bickerton}, {Bizyaev}, {Blakeslee}, {Blanton}, {Bochanski}, {Boroski}, {Brewington}, {Brinchmann}, {Brinkmann}, {Brunner}, {Budav{\'a}ri}, {Carey}, {Carliles}, {Carr}, {Castander}, {Cinabro}, {Connolly}, {Csabai}, {Cunha}, {Czarapata}, {Davenport}, {de Haas}, {Dilday}, {Doi}, {Eisenstein}, {Evans}, {Evans}, {Fan}, {Friedman}, {Frieman}, {Fukugita}, {G{\"a}nsicke}, {Gates}, {Gillespie}, {Gilmore}, {Gonzalez}, {Gonzalez}, {Grebel}, {Gunn}, {Gy{\"o}ry}, {Hall}, {Harding}, {Harris}, {Harvanek}, {Hawley}, {Hayes}, {Heckman}, {Hendry}, {Hennessy}, {Hindsley}, {Hoblitt}, {Hogan}, {Hogg}, {Holtzman}, {Hyde}, {Ichikawa}, {Ichikawa}, {Im}, {Ivezi{\'c}}, {Jester}, {Jiang}, {Johnson}, {Jorgensen}, {Juri{\'c}}, {Kent}, {Kessler}, {Kleinman},
  {Knapp}, {Konishi}, {Kron}, {Krzesinski}, {Kuropatkin}, {Lampeitl}, {Lebedeva}, {Lee}, {Lee}, {French Leger}, {L{\'e}pine}, {Li}, {Lima}, {Lin}, {Long}, {Loomis}, {Loveday}, {Lupton}, {Magnier}, {Malanushenko}, {Malanushenko}, {Mandelbaum}, {Margon}, {Marriner}, {Mart{\'\i}nez-Delgado}, {Matsubara}, {McGehee}, {McKay}, {Meiksin}, {Morrison}, {Mullally}, {Munn}, {Murphy}, {Nash}, {Nebot}, {Neilsen}, {Newberg}, {Newman}, {Nichol}, {Nicinski}, {Nieto-Santisteban}, {Nitta}, {Okamura}, {Oravetz}, {Ostriker}, {Owen}, {Padmanabhan}, {Pan}, {Park}, {Pauls}, {Peoples}, {Percival}, {Pier}, {Pope}, {Pourbaix}, {Price}, {Purger}, {Quinn}, {Raddick}, {Re Fiorentin}, {Richards}, {Richmond}, {Riess}, {Rix}, {Rockosi}, {Sako}, {Schlegel}, {Schneider}, {Scholz}, {Schreiber}, {Schwope}, {Seljak}, {Sesar}, {Sheldon}, {Shimasaku}, {Sibley}, {Simmons}, {Sivarani}, {Allyn Smith}, {Smith}, {Smol{\v{c}}i{\'c}}, {Snedden}, {Stebbins}, {Steinmetz}, {Stoughton}, {Strauss}, {SubbaRao}, {Suto}, {Szalay}, {Szapudi}, {Szkody}, {Tanaka},
  {Tegmark}, {Teodoro}, {Thakar}, {Tremonti}, {Tucker}, {Uomoto}, {Vanden Berk}, {Vandenberg}, {Vidrih}, {Vogeley}, {Voges}, {Vogt}, {Wadadekar}, {Watters}, {Weinberg}, {West}, {White}, {Wilhite}, {Wonders}, {Yanny}, {Yocum}, {York}, {Zehavi}, {Zibetti}, \& {Zucker}}]{AbazajianSDSS..Survey2009ApJS..182..543A}
{Abazajian}, K.~N., {Adelman-McCarthy}, J.~K., {Ag{\"u}eros}, M.~A., {et~al.} 2009, \bibinfo{title}{{The Seventh Data Release of the Sloan Digital Sky Survey},} \apjs, 182, 543, \dodoi{10.1088/0067-0049/182/2/543}

\bibitem[{G. {Agazie} {et~al.}(2023){Agazie}, {Anumarlapudi}, {Archibald}, {Arzoumanian}, {Baker}, {B{\'e}csy}, {Blecha}, {Brazier}, {Brook}, {Burke-Spolaor}, {Burnette}, {Case}, {Charisi}, {Chatterjee}, {Chatziioannou}, {Cheeseboro}, {Chen}, {Cohen}, {Cordes}, {Cornish}, {Crawford}, {Cromartie}, {Crowter}, {Cutler}, {Decesar}, {Degan}, {Demorest}, {Deng}, {Dolch}, {Drachler}, {Ellis}, {Ferrara}, {Fiore}, {Fonseca}, {Freedman}, {Garver-Daniels}, {Gentile}, {Gersbach}, {Glaser}, {Good}, {G{\"u}ltekin}, {Hazboun}, {Hourihane}, {Islo}, {Jennings}, {Johnson}, {Jones}, {Kaiser}, {Kaplan}, {Kelley}, {Kerr}, {Key}, {Klein}, {Laal}, {Lam}, {Lamb}, {Lazio}, {Lewandowska}, {Littenberg}, {Liu}, {Lommen}, {Lorimer}, {Luo}, {Lynch}, {Ma}, {Madison}, {Mattson}, {McEwen}, {McKee}, {McLaughlin}, {McMann}, {Meyers}, {Meyers}, {Mingarelli}, {Mitridate}, {Natarajan}, {Ng}, {Nice}, {Ocker}, {Olum}, {Pennucci}, {Perera}, {Petrov}, {Pol}, {Radovan}, {Ransom}, {Ray}, {Romano}, {Sardesai}, {Schmiedekamp}, {Schmiedekamp}, {Schmitz},
  {Schult}, {Shapiro-Albert}, {Siemens}, {Simon}, {Siwek}, {Stairs}, {Stinebring}, {Stovall}, {Sun}, {Susobhanan}, {Swiggum}, {Taylor}, {Taylor}, {Turner}, {Unal}, {Vallisneri}, {van Haasteren}, {Vigeland}, {Wahl}, {Wang}, {Witt}, {Young}, \& {Nanograv Collaboration}}]{agazie.etal.2023}
{Agazie}, G., {Anumarlapudi}, A., {Archibald}, A.~M., {et~al.} 2023, \bibinfo{title}{{The NANOGrav 15 yr Data Set: Evidence for a Gravitational-wave Background},} \apjl, 951, L8, \dodoi{10.3847/2041-8213/acdac6}

\bibitem[{C.~P. {Ahn} {et~al.}(2012){Ahn}, {Alexandroff}, {Allende Prieto}, {Anderson}, {Anderton}, {Andrews}, {Aubourg}, {Bailey}, {Balbinot}, {Barnes}, {Bautista}, {Beers}, {Beifiori}, {Berlind}, {Bhardwaj}, {Bizyaev}, {Blake}, {Blanton}, {Blomqvist}, {Bochanski}, {Bolton}, {Borde}, {Bovy}, {Brandt}, {Brinkmann}, {Brown}, {Brownstein}, {Bundy}, {Busca}, {Carithers}, {Carnero}, {Carr}, {Casetti-Dinescu}, {Chen}, {Chiappini}, {Comparat}, {Connolly}, {Crepp}, {Cristiani}, {Croft}, {Cuesta}, {da Costa}, {Davenport}, {Dawson}, {de Putter}, {De Lee}, {Delubac}, {Dhital}, {Ealet}, {Ebelke}, {Edmondson}, {Eisenstein}, {Escoffier}, {Esposito}, {Evans}, {Fan}, {Femen{\'\i}a Castell{\'a}}, {Fern{\'a}ndez Alvar}, {Ferreira}, {Filiz Ak}, {Finley}, {Fleming}, {Font-Ribera}, {Frinchaboy}, {Garc{\'\i}a-Hern{\'a}ndez}, {Garc{\'\i}a P{\'e}rez}, {Ge}, {G{\'e}nova-Santos}, {Gillespie}, {Girardi}, {Gonz{\'a}lez Hern{\'a}ndez}, {Grebel}, {Gunn}, {Guo}, {Haggard}, {Hamilton}, {Harris}, {Hawley}, {Hearty}, {Ho}, {Hogg},
  {Holtzman}, {Honscheid}, {Huehnerhoff}, {Ivans}, {Ivezi{\'c}}, {Jacobson}, {Jiang}, {Johansson}, {Johnson}, {Kauffmann}, {Kirkby}, {Kirkpatrick}, {Klaene}, {Knapp}, {Kneib}, {Le Goff}, {Leauthaud}, {Lee}, {Lee}, {Long}, {Loomis}, {Lucatello}, {Lundgren}, {Lupton}, {Ma}, {Ma}, {MacDonald}, {Mack}, {Mahadevan}, {Maia}, {Majewski}, {Makler}, {Malanushenko}, {Malanushenko}, {Manchado}, {Mandelbaum}, {Manera}, {Maraston}, {Margala}, {Martell}, {McBride}, {McGreer}, {McMahon}, {M{\'e}nard}, {Meszaros}, {Miralda-Escud{\'e}}, {Montero-Dorta}, {Montesano}, {Morrison}, {Muna}, {Munn}, {Murayama}, {Myers}, {Neto}, {Nguyen}, {Nichol}, {Nidever}, {Noterdaeme}, {Nuza}, {Ogando}, {Olmstead}, {Oravetz}, {Owen}, {Padmanabhan}, {Palanque-Delabrouille}, {Pan}, {Parejko}, {Parihar}, {P{\^a}ris}, {Pattarakijwanich}, {Pepper}, {Percival}, {P{\'e}rez-Fournon}, {P{\'e}rez-R{\`a}fols}, {Petitjean}, {Pforr}, {Pieri}, {Pinsonneault}, {Porto de Mello}, {Prada}, {Price-Whelan}, {Raddick}, {Rebolo}, {Rich}, {Richards}, {Robin},
  {Rocha-Pinto}, {Rockosi}, {Roe}, {Ross}, {Ross}, {Rossi}, {Rubi{\~n}o-Martin}, {Samushia}, {Sanchez Almeida}, {S{\'a}nchez}, {Santiago}, {Sayres}, {Schlegel}, {Schlesinger}, {Schmidt}, {Schneider}, {Schultheis}, {Schwope}, {Sc{\'o}ccola}, {Seljak}, {Sheldon}, {Shen}, {Shu}, {Simmerer}, {Simmons}, {Skibba}, {Skrutskie}, {Slosar}, {Sobreira}, {Sobeck}, {Stassun}, {Steele}, {Steinmetz}, {Strauss}, {Streblyanska}, {Suzuki}, {Swanson}, {Tal}, {Thakar}, {Thomas}, {Thompson}, {Tinker}, {Tojeiro}, {Tremonti}, {Vargas Maga{\~n}a}, {Verde}, {Viel}, {Vikas}, {Vogt}, {Wake}, {Wang}, {Weaver}, {Weinberg}, {Weiner}, {West}, {White}, {Wilson}, {Wisniewski}, {Wood-Vasey}, {Yanny}, {Y{\`e}che}, {York}, {Zamora}, {Zasowski}, {Zehavi}, {Zhao}, {Zheng}, {Zhu}, \& {Zinn}}]{AhnBOSS..SDSS..Survey2012ApJS..203...21A}
{Ahn}, C.~P., {Alexandroff}, R., {Allende Prieto}, C., {et~al.} 2012, \bibinfo{title}{{The Ninth Data Release of the Sloan Digital Sky Survey: First Spectroscopic Data from the SDSS-III Baryon Oscillation Spectroscopic Survey},} \apjs, 203, 21, \dodoi{10.1088/0067-0049/203/2/21}

\bibitem[{A.~S.~R. {Antas} {et~al.}(2024){Antas}, {Caproni}, {Machado}, {Lagan{\'a}}, \& {Souza}}]{Antas2024}
{Antas}, A.~S.~R., {Caproni}, A., {Machado}, R.~E.~G., {Lagan{\'a}}, T.~F., \& {Souza}, G.~S. 2024, \bibinfo{title}{{Orbital motion of NGC 6166 (3C 338) and its impact on the jet morphology at kiloparsec scales},} \mnras, 533, 1341, \dodoi{10.1093/mnras/stae1846}

\bibitem[{K. {Bansal} {et~al.}(2017){Bansal}, {Taylor}, {Peck}, {Zavala}, \& {Romani}}]{BansalEtal2017ApJ...843...14B}
{Bansal}, K., {Taylor}, G.~B., {Peck}, A.~B., {Zavala}, R.~T., \& {Romani}, R.~W. 2017, \bibinfo{title}{{Constraining the Orbit of the Supermassive Black Hole Binary 0402+379},} \apj, 843, 14, \dodoi{10.3847/1538-4357/aa74e1}

\bibitem[{J.~M. {Bardeen} \& J.~A. {Petterson}(1975){Bardeen} \& {Petterson}}]{BardeenPetterson1975ApJ...195L..65B}
{Bardeen}, J.~M., \& {Petterson}, J.~A. 1975, \bibinfo{title}{{The Lense-Thirring Effect and Accretion Disks around Kerr Black Holes},} \apjl, 195, L65, \dodoi{10.1086/181711}

\bibitem[{M.~C. {Begelman} {et~al.}(1980){Begelman}, {Blandford}, \& {Rees}}]{BegelmanEtal1980Natur.287..307B}
{Begelman}, M.~C., {Blandford}, R.~D., \& {Rees}, M.~J. 1980, \bibinfo{title}{{Massive black hole binaries in active galactic nuclei},} \nat, 287, 307, \dodoi{10.1038/287307a0}

\bibitem[{R.~D. {Blandford} \& D.~G. {Payne}(1982){Blandford} \& {Payne}}]{BP1982MNRAS.199..883B}
{Blandford}, R.~D., \& {Payne}, D.~G. 1982, \bibinfo{title}{{Hydromagnetic flows from accretion disks and the production of radio jets.},} \mnras, 199, 883, \dodoi{10.1093/mnras/199.4.883}

\bibitem[{R.~D. {Blandford} \& R.~L. {Znajek}(1977){Blandford} \& {Znajek}}]{BZ1977MNRAS.179..433B}
{Blandford}, R.~D., \& {Znajek}, R.~L. 1977, \bibinfo{title}{{Electromagnetic extraction of energy from Kerr black holes.},} \mnras, 179, 433, \dodoi{10.1093/mnras/179.3.433}

\bibitem[{P. {Breiding} {et~al.}(2022){Breiding}, {Burke-Spolaor}, {An}, {Bansal}, {Mohan}, {Taylor}, \& {Zhang}}]{BreidingEtal2022ApJ...933..143B}
{Breiding}, P., {Burke-Spolaor}, S., {An}, T., {et~al.} 2022, \bibinfo{title}{{Deep Very Long Baseline Interferometry Observations Challenge Previous Evidence of a Binary Supermassive Black Hole Residing in Seyfert Galaxy NGC 7674},} \apj, 933, 143, \dodoi{10.3847/1538-4357/ac7466}

\bibitem[{L. {Bruno} {et~al.}(2019){Bruno}, {Gitti}, {Zanichelli}, \& {Gregorini}}]{Bruno2019}
{Bruno}, L., {Gitti}, M., {Zanichelli}, A., \& {Gregorini}, L. 2019, \bibinfo{title}{{Multifrequency JVLA observations of the X-shaped radio galaxy in Abell 3670},} \aap, 631, A173, \dodoi{10.1051/0004-6361/201936240}

\bibitem[{S. {Burke-Spolaor} {et~al.}(2014){Burke-Spolaor}, {Brazier}, {Chatterjee}, {Comerford}, {Cordes}, {Lazio}, {Liu}, \& {Shen}}]{Burke-Spolaor2014arXiv1402.0548B}
{Burke-Spolaor}, S., {Brazier}, A., {Chatterjee}, S., {et~al.} 2014, \bibinfo{title}{{A hunt for dual radio active galactic nuclei in the VLASS},} arXiv e-prints, arXiv:1402.0548, \dodoi{10.48550/arXiv.1402.0548}

\bibitem[{J.~O. {Burns} \& T.~J. {Balonek}(1982){Burns} \& {Balonek}}]{BurnsBalonek1982ApJ...263..546B}
{Burns}, J.~O., \& {Balonek}, T.~J. 1982, \bibinfo{title}{{The curvature of radio jets and tails in the intracluster media of Abell 1446 and 2220.},} \apj, 263, 546, \dodoi{10.1086/160525}

\bibitem[{Y. {Cui} {et~al.}(2023){Cui}, {Hada}, {Kawashima}, {Kino}, {Lin}, {Mizuno}, {Ro}, {Honma}, {Yi}, {Yu}, {Park}, {Jiang}, {Shen}, {Kravchenko}, {Algaba}, {Cheng}, {Cho}, {Giovannini}, {Giroletti}, {Jung}, {Lu}, {Niinuma}, {Oh}, {Ohsuga}, {Sawada-Satoh}, {Sohn}, {Takahashi}, {Takamura}, {Tazaki}, {Trippe}, {Wajima}, {Akiyama}, {An}, {Asada}, {Buttaccio}, {Byun}, {Cui}, {Hagiwara}, {Hirota}, {Hodgson}, {Kawaguchi}, {Kim}, {Lee}, {Lee}, {Lee}, {Maccaferri}, {Melis}, {Melnikov}, {Migoni}, {Oh}, {Sugiyama}, {Wang}, {Zhang}, {Chen}, {Hwang}, {Jung}, {Kim}, {Kim}, {Kobayashi}, {Li}, {Li}, {Li}, {Liu}, {Liu}, {Liu}, {Oh}, {Oyama}, {Roh}, {Wang}, {Wang}, {Wang}, {Xia}, {Yan}, {Yeom}, {Yonekura}, {Yuan}, {Zhang}, {Zhao}, \& {Zhong}}]{CuiEtAl2023Natur.621..711C}
{Cui}, Y., {Hada}, K., {Kawashima}, T., {et~al.} 2023, \bibinfo{title}{{Precessing jet nozzle connecting to a spinning black hole in M87},} \nat, 621, 711, \dodoi{10.1038/s41586-023-06479-6}

\bibitem[{A. {De Rosa} {et~al.}(2019){De Rosa}, {Vignali}, {Bogdanovi{\'c}}, {Capelo}, {Charisi}, {Dotti}, {Husemann}, {Lusso}, {Mayer}, {Paragi}, {Runnoe}, {Sesana}, {Steinborn}, {Bianchi}, {Colpi}, {del Valle}, {Frey}, {Gab{\'a}nyi}, {Giustini}, {Guainazzi}, {Haiman}, {Herrera Ruiz}, {Herrero-Illana}, {Iwasawa}, {Komossa}, {Lena}, {Loiseau}, {Perez-Torres}, {Piconcelli}, \& {Volonteri}}]{DeRosaEtalReview2019NewAR..8601525D}
{De Rosa}, A., {Vignali}, C., {Bogdanovi{\'c}}, T., {et~al.} 2019, \bibinfo{title}{{The quest for dual and binary supermassive black holes: A multi-messenger view},} \nar, 86, 101525, \dodoi{10.1016/j.newar.2020.101525}

\bibitem[{R.~P. {Dubey} {et~al.}(2023){Dubey}, {Fendt}, \& {Vaidya}}]{DubeyEtal2023ApJ...952....1D}
{Dubey}, R.~P., {Fendt}, C., \& {Vaidya}, B. 2023, \bibinfo{title}{{Particles in Relativistic MHD Jets. I. Role of Jet Dynamics in Particle Acceleration},} \apj, 952, 1, \dodoi{10.3847/1538-4357/ace0bf}

\bibitem[{R.~J.~H. {Dunn} {et~al.}(2006){Dunn}, {Fabian}, \& {Sanders}}]{DunnEtAl2006MNRAS.366..758D}
{Dunn}, R.~J.~H., {Fabian}, A.~C., \& {Sanders}, J.~S. 2006, \bibinfo{title}{{Precession of the super-massive black hole in NGC 1275 (3C 84)?},} \mnras, 366, 758, \dodoi{10.1111/j.1365-2966.2005.09928.x}

\bibitem[{D. {Falceta-Gon{\c{c}}alves} {et~al.}(2010){Falceta-Gon{\c{c}}alves}, {Caproni}, {Abraham}, {Teixeira}, \& {de Gouveia Dal Pino}}]{Falceta-GoncalvesEtAl2010ApJ...713L..74F}
{Falceta-Gon{\c{c}}alves}, D., {Caproni}, A., {Abraham}, Z., {Teixeira}, D.~M., \& {de Gouveia Dal Pino}, E.~M. 2010, \bibinfo{title}{{Precessing Jets and X-ray Bubbles from NGC 1275 (3C 84) in the Perseus Galaxy Cluster: A View from Three-dimensional Numerical Simulations},} \apjl, 713, L74, \dodoi{10.1088/2041-8205/713/1/L74}

\bibitem[{R.~P. {Fender} {et~al.}(2004){Fender}, {Belloni}, \& {Gallo}}]{FenderEtal2004MNRAS.355.1105F}
{Fender}, R.~P., {Belloni}, T.~M., \& {Gallo}, E. 2004, \bibinfo{title}{{Towards a unified model for black hole X-ray binary jets},} \mnras, 355, 1105, \dodoi{10.1111/j.1365-2966.2004.08384.x}

\bibitem[{P.~C. {Fragile} {et~al.}(2007){Fragile}, {Blaes}, {Anninos}, \& {Salmonson}}]{Fragile2007}
{Fragile}, P.~C., {Blaes}, O.~M., {Anninos}, P., \& {Salmonson}, J.~D. 2007, \bibinfo{title}{{Global General Relativistic Magnetohydrodynamic Simulation of a Tilted Black Hole Accretion Disk},} \apj, 668, 417, \dodoi{10.1086/521092}

\bibitem[{H. {Fu} {et~al.}(2012){Fu}, {Yan}, {Myers}, {Stockton}, {Djorgovski}, {Aldering}, \& {Rich}}]{Fu2012}
{Fu}, H., {Yan}, L., {Myers}, A.~D., {et~al.} 2012, \bibinfo{title}{{The Nature of Double-peaked [O III] Active Galactic Nuclei},} \apj, 745, 67, \dodoi{10.1088/0004-637X/745/1/67}

\bibitem[{M. {Gaspari} {et~al.}(2020){Gaspari}, {Tombesi}, \& {Cappi}}]{Gaspari2020}
{Gaspari}, M., {Tombesi}, F., \& {Cappi}, M. 2020, \bibinfo{title}{{Linking macro-, meso- and microscales in multiphase AGN feeding and feedback},} Nature Astronomy, 4, 10, \dodoi{10.1038/s41550-019-0970-1}

\bibitem[{G. Giri {et~al.}(2024)Giri, Bagchi, Thorat, Deane, Delhaize, \& Saikia}]{Giri2024}
Giri, G., Bagchi, J., Thorat, K., {et~al.} 2024, \bibinfo{title}{Probing the Formation of Megaparsec-scale Giant Radio Galaxies (I): Dynamical Insights from MHD Simulations,} arXiv preprint arXiv:2411.10864

\bibitem[{G. {Giri} {et~al.}(2022{\natexlab{a}}){Giri}, {Dubey}, {Rubinur}, {Vaidya}, \& {Kharb}}]{GiriEtal2022MNRAS.514.5625G}
{Giri}, G., {Dubey}, R.~P., {Rubinur}, K., {Vaidya}, B., \& {Kharb}, P. 2022{\natexlab{a}}, \bibinfo{title}{{Dynamical modelling and emission signatures of a candidate dual AGN with precessing radio jets},} \mnras, 514, 5625, \dodoi{10.1093/mnras/stac1628}

\bibitem[{G. {Giri} {et~al.}(2023){Giri}, {Vaidya}, \& {Fendt}}]{Giri2023}
{Giri}, G., {Vaidya}, B., \& {Fendt}, C. 2023, \bibinfo{title}{{Deciphering the Morphological Origins of X-shaped Radio Galaxies: Numerical Modeling of Backflow versus Jet Reorientation},} \apjs, 268, 49, \dodoi{10.3847/1538-4365/acebca}

\bibitem[{G. {Giri} {et~al.}(2022{\natexlab{b}}){Giri}, {Vaidya}, {Rossi}, {Bodo}, {Mukherjee}, \& {Mignone}}]{GiriEtal2022A&A...662A...5G}
{Giri}, G., {Vaidya}, B., {Rossi}, P., {et~al.} 2022{\natexlab{b}}, \bibinfo{title}{{Modelling X-shaped radio galaxies: Dynamical and emission signatures from the Back-flow model},} \aap, 662, A5, \dodoi{10.1051/0004-6361/202142546}

\bibitem[{ {Gopal-Krishna} {et~al.}(2003){Gopal-Krishna}, {Biermann}, \& {Wiita}}]{GoaplEtal2003ApJ...594L.103G}
{Gopal-Krishna}, {Biermann}, P.~L., \& {Wiita}, P.~J. 2003, \bibinfo{title}{{The Origin of X-shaped Radio Galaxies: Clues from the Z-symmetric Secondary Lobes},} \apjl, 594, L103, \dodoi{10.1086/378766}

\bibitem[{ {Gopal-Krishna} {et~al.}(2022){Gopal-Krishna}, {Joshi}, {Patra}, {Yang}, {Ho}, {Wiita}, \& {Omar}}]{Gopal-KrishnaEtal2022MNRAS.514L..36G}
{Gopal-Krishna}, {Joshi}, R., {Patra}, D., {et~al.} 2022, \bibinfo{title}{{The twin radio galaxy TRG J104454+354055},} \mnras, 514, L36, \dodoi{10.1093/mnrasl/slac055}

\bibitem[{ {Gopal-Krishna} \& P.~J. {Wiita}(1987){Gopal-Krishna} \& {Wiita}}]{GopalWiita1987MNRAS.226..531G}
{Gopal-Krishna}, \& {Wiita}, P.~J. 1987, \bibinfo{title}{{The expansion and cosmological evolution of powerful radio sources},} \mnras, 226, 531, \dodoi{10.1093/mnras/226.3.531}

\bibitem[{A.~C. {Gower} {et~al.}(1982){Gower}, {Gregory}, {Unruh}, \& {Hutchings}}]{GowerEtAl1982ApJ...262..478G}
{Gower}, A.~C., {Gregory}, P.~C., {Unruh}, W.~G., \& {Hutchings}, J.~B. 1982, \bibinfo{title}{{Relativistic precessing jets in quasars and radio galaxies : models to fit high resolution data.},} \apj, 262, 478, \dodoi{10.1086/160442}

\bibitem[{M.~J. {Graham} {et~al.}(2015){Graham}, {Djorgovski}, {Stern}, {Glikman}, {Drake}, {Mahabal}, {Donalek}, {Larson}, \& {Christensen}}]{GrahamEtal2015Natur.518...74G}
{Graham}, M.~J., {Djorgovski}, S.~G., {Stern}, D., {et~al.} 2015, \bibinfo{title}{{A possible close supermassive black-hole binary in a quasar with optical periodicity},} \nat, 518, 74, \dodoi{10.1038/nature14143}

\bibitem[{D. {Guidetti} {et~al.}(2008){Guidetti}, {Murgia}, {Govoni}, {Parma}, {Gregorini}, {de Ruiter}, {Cameron}, \& {Fanti}}]{GuidettiEtal2008A&A...483..699G}
{Guidetti}, D., {Murgia}, M., {Govoni}, F., {et~al.} 2008, \bibinfo{title}{{The intracluster magnetic field power spectrum in Abell 2382},} \aap, 483, 699, \dodoi{10.1051/0004-6361:20078576}

\bibitem[{M.~J. {Hardcastle}(2018){Hardcastle}}]{Hardcastle2018}
{Hardcastle}, M.~J. 2018, \bibinfo{title}{{A simulation-based analytic model of radio galaxies},} \mnras, 475, 2768, \dodoi{10.1093/mnras/stx3358}

\bibitem[{M.~J. {Hardcastle} \& M.~G.~H. {Krause}(2013){Hardcastle} \& {Krause}}]{Hardcastle2013}
{Hardcastle}, M.~J., \& {Krause}, M.~G.~H. 2013, \bibinfo{title}{{Numerical modelling of the lobes of radio galaxies in cluster environments},} \mnras, 430, 174, \dodoi{10.1093/mnras/sts564}

\bibitem[{M.~J. {Hardcastle} \& M.~G.~H. {Krause}(2014){Hardcastle} \& {Krause}}]{HardcastleHrause2014MNRAS.443.1482H}
{Hardcastle}, M.~J., \& {Krause}, M.~G.~H. 2014, \bibinfo{title}{{Numerical modelling of the lobes of radio galaxies in cluster environments - II. Magnetic field configuration and observability},} \mnras, 443, 1482, \dodoi{10.1093/mnras/stu1229}

\bibitem[{R.~M. {Hjellming} \& K.~J. {Johnston}(1981){Hjellming} \& {Johnston}}]{HjellmingJohnston1981ApJ...246L.141H}
{Hjellming}, R.~M., \& {Johnston}, K.~J. 1981, \bibinfo{title}{{An analysis of the proper motions of SS 433 radio jets.},} \apjl, 246, L141, \dodoi{10.1086/183571}

\bibitem[{P.~F. {Hopkins} {et~al.}(2012){Hopkins}, {Hernquist}, {Hayward}, \& {Narayanan}}]{Hopkins2012}
{Hopkins}, P.~F., {Hernquist}, L., {Hayward}, C.~C., \& {Narayanan}, D. 2012, \bibinfo{title}{{Why are active galactic nuclei and host galaxies misaligned?},} \mnras, 425, 1121, \dodoi{10.1111/j.1365-2966.2012.21449.x}

\bibitem[{M.~E. {Jarvis} {et~al.}(2019){Jarvis}, {Harrison}, {Thomson}, {Circosta}, {Mainieri}, {Alexander}, {Edge}, {Lansbury}, {Molyneux}, \& {Mullaney}}]{JarvisEtal2019MNRAS.485.2710J}
{Jarvis}, M.~E., {Harrison}, C.~M., {Thomson}, A.~P., {et~al.} 2019, \bibinfo{title}{{Prevalence of radio jets associated with galactic outflows and feedback from quasars},} \mnras, 485, 2710, \dodoi{10.1093/mnras/stz556}

\bibitem[{R. {Joshi} {et~al.}(2019){Joshi}, {Krishna}, {Yang}, {Shi}, {Yu}, {Wiita}, {Ho}, {Wu}, {An}, {Wang}, {Subramanian}, \& {Yesuf}}]{JoshiEtal2019ApJ...887..266J}
{Joshi}, R., {Krishna}, G., {Yang}, X., {et~al.} 2019, \bibinfo{title}{{X-shaped Radio Galaxies: Optical Properties, Large-scale Environment, and Relationship to Radio Structure},} \apj, 887, 266, \dodoi{10.3847/1538-4357/ab536f}

\bibitem[{P. {Kharb} {et~al.}(2017){Kharb}, {Lal}, \& {Merritt}}]{kharb.lal.etal.2017}
{Kharb}, P., {Lal}, D.~V., \& {Merritt}, D. 2017, \bibinfo{title}{{A candidate sub-parsec binary black hole in the Seyfert galaxy NGC 7674},} Nature Astronomy, 1, 727, \dodoi{10.1038/s41550-017-0256-4}

\bibitem[{P. {Kharb} {et~al.}(2019){Kharb}, {Vaddi}, {Sebastian}, {Subramanian}, {Das}, \& {Paragi}}]{Kharb2019}
{Kharb}, P., {Vaddi}, S., {Sebastian}, B., {et~al.} 2019, \bibinfo{title}{{A Curved 150 pc Long Jet in the Double-peaked Emission-line AGN KISSR 434},} \apj, 871, 249, \dodoi{10.3847/1538-4357/aafad7}

\bibitem[{J. {Kormendy} \& L.~C. {Ho}(2013){Kormendy} \& {Ho}}]{KormendyHo2013ARA&A..51..511K}
{Kormendy}, J., \& {Ho}, L.~C. 2013, \bibinfo{title}{{Coevolution (Or Not) of Supermassive Black Holes and Host Galaxies},} \araa, 51, 511, \dodoi{10.1146/annurev-astro-082708-101811}

\bibitem[{J. {Kormendy} \& D. {Richstone}(1995){Kormendy} \& {Richstone}}]{KormendyRichstone1995ARA&A..33..581K}
{Kormendy}, J., \& {Richstone}, D. 1995, \bibinfo{title}{{Inward Bound---The Search For Supermassive Black Holes In Galactic Nuclei},} \araa, 33, 581, \dodoi{10.1146/annurev.aa.33.090195.003053}

\bibitem[{D.~V. {Lal} \& A.~P. {Rao}(2005){Lal} \& {Rao}}]{Lal2005}
{Lal}, D.~V., \& {Rao}, A.~P. 2005, \bibinfo{title}{{3C 223.1: A source with unusual spectral properties},} \mnras, 356, 232, \dodoi{10.1111/j.1365-2966.2004.08442.x}

\bibitem[{D.~V. {Lal} {et~al.}(2019){Lal}, {Sebastian}, {Cheung}, \& {Pramesh Rao}}]{Lal2019}
{Lal}, D.~V., {Sebastian}, B., {Cheung}, C.~C., \& {Pramesh Rao}, A. 2019, \bibinfo{title}{{GMRT Low-frequency Imaging of an Extended Sample of X-shaped Radio Galaxies},} \aj, 157, 195, \dodoi{10.3847/1538-3881/ab1419}

\bibitem[{M. {Liska} {et~al.}(2018){Liska}, {Hesp}, {Tchekhovskoy}, {Ingram}, {van der Klis}, \& {Markoff}}]{LiskaEtAl2018MNRAS.474L..81L}
{Liska}, M., {Hesp}, C., {Tchekhovskoy}, A., {et~al.} 2018, \bibinfo{title}{{Formation of precessing jets by tilted black hole discs in 3D general relativistic MHD simulations},} \mnras, 474, L81, \dodoi{10.1093/mnrasl/slx174}

\bibitem[{T. {Liu} {et~al.}(2015){Liu}, {Gezari}, {Heinis}, {Magnier}, {Burgett}, {Chambers}, {Flewelling}, {Huber}, {Hodapp}, {Kaiser}, {Kudritzki}, {Tonry}, {Wainscoat}, \& {Waters}}]{LiuEtal2015ApJ...803L..16L}
{Liu}, T., {Gezari}, S., {Heinis}, S., {et~al.} 2015, \bibinfo{title}{{A Periodically Varying Luminous Quasar at z = 2 from the Pan-STARRS1 Medium Deep Survey: A Candidate Supermassive Black Hole Binary in the Gravitational Wave-driven Regime},} \apjl, 803, L16, \dodoi{10.1088/2041-8205/803/2/L16}

\bibitem[{J. {Magorrian} {et~al.}(1998){Magorrian}, {Tremaine}, {Richstone}, {Bender}, {Bower}, {Dressler}, {Faber}, {Gebhardt}, {Green}, {Grillmair}, {Kormendy}, \& {Lauer}}]{MagorrianEtal1998AJ....115.2285M}
{Magorrian}, J., {Tremaine}, S., {Richstone}, D., {et~al.} 1998, \bibinfo{title}{{The Demography of Massive Dark Objects in Galaxy Centers},} \aj, 115, 2285, \dodoi{10.1086/300353}

\bibitem[{A. {Mignone} {et~al.}(2007){Mignone}, {Bodo}, {Massaglia}, {Matsakos}, {Tesileanu}, {Zanni}, \& {Ferrari}}]{Mignone2007}
{Mignone}, A., {Bodo}, G., {Massaglia}, S., {et~al.} 2007, \bibinfo{title}{{PLUTO: A Numerical Code for Computational Astrophysics},} \apjs, 170, 228, \dodoi{10.1086/513316}

\bibitem[{A.~J. {Mioduszewski} {et~al.}(1997){Mioduszewski}, {Hughes}, \& {Duncan}}]{MioduszewskiEtal1997ApJ...476..649M}
{Mioduszewski}, A.~J., {Hughes}, P.~A., \& {Duncan}, G.~C. 1997, \bibinfo{title}{{Simulated VLBI Images from Relativistic Hydrodynamic Jet Models},} \apj, 476, 649, \dodoi{10.1086/303652}

\bibitem[{S.~M. {Molnar} {et~al.}(2017){Molnar}, {Schive}, {Birkinshaw}, {Chiueh}, {Musoke}, \& {Young}}]{MolnarEtal2017ApJ...835...57M}
{Molnar}, S.~M., {Schive}, H.~Y., {Birkinshaw}, M., {et~al.} 2017, \bibinfo{title}{{Hydrodynamical Simulations of Colliding Jets: Modeling 3C 75},} \apj, 835, 57, \dodoi{10.3847/1538-4357/835/1/57}

\bibitem[{S. {Mondal} {et~al.}(2022{\natexlab{a}}){Mondal}, {Keshet}, {Sarkar}, \& {Gurwich}}]{MondalEtal2022MNRAS.514.2581M}
{Mondal}, S., {Keshet}, U., {Sarkar}, K.~C., \& {Gurwich}, I. 2022{\natexlab{a}}, \bibinfo{title}{{Fermi bubbles: the collimated outburst needed to explain forward-shock edges},} \mnras, 514, 2581, \dodoi{10.1093/mnras/stac1084}

\bibitem[{S. {Mondal} {et~al.}(2022{\natexlab{b}}){Mondal}, {Rani}, {Stalin}, {Chakrabarti}, \& {Rakshit}}]{MondalEtal2022AA...663A.178M}
{Mondal}, S., {Rani}, P., {Stalin}, C.~S., {Chakrabarti}, S.~K., \& {Rakshit}, S. 2022{\natexlab{b}}, \bibinfo{title}{{Flux and spectral variability of Mrk 421 during its moderate activity state using NuSTAR: Possible accretion disc contribution?},} \aap, 663, A178, \dodoi{10.1051/0004-6361/202141990}

\bibitem[{M. {Murgia} {et~al.}(2001){Murgia}, {Parma}, {de Ruiter}, {Bondi}, {Ekers}, {Fanti}, \& {Fomalont}}]{Murgia2001}
{Murgia}, M., {Parma}, P., {de Ruiter}, H.~R., {et~al.} 2001, \bibinfo{title}{{A multi-frequency study of the radio galaxy NGC 326. I. The data},} \aap, 380, 102, \dodoi{10.1051/0004-6361:20011436}

\bibitem[{G. {Musoke} {et~al.}(2020{\natexlab{a}}){Musoke}, {Young}, {Molnar}, \& {Birkinshaw}}]{Musoke2020}
{Musoke}, G., {Young}, A.~J., {Molnar}, S.~M., \& {Birkinshaw}, M. 2020{\natexlab{a}}, \bibinfo{title}{{Numerical simulations of colliding jets in an external wind: application to 3C 75},} \mnras, 494, 5207, \dodoi{10.1093/mnras/staa1071}

\bibitem[{G. {Musoke} {et~al.}(2020{\natexlab{b}}){Musoke}, {Young}, {Molnar}, \& {Birkinshaw}}]{MusokeEtal2020MNRAS.494.5207M}
{Musoke}, G., {Young}, A.~J., {Molnar}, S.~M., \& {Birkinshaw}, M. 2020{\natexlab{b}}, \bibinfo{title}{{Numerical simulations of colliding jets in an external wind: application to 3C 75},} \mnras, 494, 5207, \dodoi{10.1093/mnras/staa1071}

\bibitem[{S. {Nandi} {et~al.}(2017){Nandi}, {Jamrozy}, {Roy}, {Larsson}, {Saikia}, {Baes}, \& {Singh}}]{NandiEtal2017MNRAS.467L..56N}
{Nandi}, S., {Jamrozy}, M., {Roy}, R., {et~al.} 2017, \bibinfo{title}{{Tale of J1328+2752: a misaligned double-double radio galaxy hosted by a binary black hole?},} \mnras, 467, L56, \dodoi{10.1093/mnrasl/slw256}

\bibitem[{R. {Narayan} \& I. {Yi}(1994){Narayan} \& {Yi}}]{NarayanYi1994ApJ...428L..13N}
{Narayan}, R., \& {Yi}, I. 1994, \bibinfo{title}{{Advection-dominated Accretion: A Self-similar Solution},} \apjl, 428, L13, \dodoi{10.1086/187381}

\bibitem[{M.~A. {Nawaz} {et~al.}(2016){Nawaz}, {Bicknell}, {Wagner}, {Sutherland}, \& {McNamara}}]{NawazEtAl2016MNRAS.458..802N}
{Nawaz}, M.~A., {Bicknell}, G.~V., {Wagner}, A.~Y., {Sutherland}, R.~S., \& {McNamara}, B.~R. 2016, \bibinfo{title}{{Jet-intracluster medium interaction in Hydra A - II. The effect of jet precession},} \mnras, 458, 802, \dodoi{10.1093/mnras/stw330}

\bibitem[{C. {Nolting} {et~al.}(2023){Nolting}, {Ball}, \& {Nguyen}}]{NoltingEtAl2023ApJ...948...25N}
{Nolting}, C., {Ball}, J., \& {Nguyen}, T.~M. 2023, \bibinfo{title}{{Simulations of Precessing Jets and the Formation of X-shaped Radio Galaxies},} \apj, 948, 25, \dodoi{10.3847/1538-4357/acc652}

\bibitem[{F.~N. {Owen} {et~al.}(1985){Owen}, {O'Dea}, {Inoue}, \& {Eilek}}]{OwenEtal1985ApJ...294L..85O}
{Owen}, F.~N., {O'Dea}, C.~P., {Inoue}, M., \& {Eilek}, J.~A. 1985, \bibinfo{title}{{VLA observations of the multiple jet galaxy 3C 75.},} \apjl, 294, L85, \dodoi{10.1086/184514}

\bibitem[{P. {Parma} {et~al.}(1991){Parma}, {de Ruiter}, \& {Cameron}}]{ParmaEtal1991AJ....102.1960P}
{Parma}, P., {de Ruiter}, H.~R., \& {Cameron}, R.~A. 1991, \bibinfo{title}{{Very Large Array Observations of Radio-Selected Dumbbell Galaxies},} \aj, 102, 1960, \dodoi{10.1086/116018}

\bibitem[{D. {Patra} {et~al.}(2023){Patra}, {Joshi}, \& {Gopal-Krishna}}]{Patra2023}
{Patra}, D., {Joshi}, R., \& {Gopal-Krishna}. 2023, \bibinfo{title}{{Spectral index variation across X-shaped radio galaxies},} \mnras, 524, 3270, \dodoi{10.1093/mnras/stad1997}

\bibitem[{F. {Pizzolato} \& N. {Soker}(2005){Pizzolato} \& {Soker}}]{PizzolatoSoker2005AdSpR..36..762P}
{Pizzolato}, F., \& {Soker}, N. 2005, \bibinfo{title}{{Binary black holes at the core of galaxy clusters},} Advances in Space Research, 36, 762, \dodoi{10.1016/j.asr.2005.02.020}

\bibitem[{J.~E. {Pringle}(1992){Pringle}}]{Pringle1992MNRAS.258..811P}
{Pringle}, J.~E. 1992, \bibinfo{title}{{A simple approach to the evolution of twisted accretion discs},} \mnras, 258, 811, \dodoi{10.1093/mnras/258.4.811}

\bibitem[{C. {Rodriguez} {et~al.}(2006){Rodriguez}, {Taylor}, {Zavala}, {Peck}, {Pollack}, \& {Romani}}]{RodriguezEtal2006ApJ...646...49R}
{Rodriguez}, C., {Taylor}, G.~B., {Zavala}, R.~T., {et~al.} 2006, \bibinfo{title}{{A Compact Supermassive Binary Black Hole System},} \apj, 646, 49, \dodoi{10.1086/504825}

\bibitem[{P. {Rossi} {et~al.}(2017){Rossi}, {Bodo}, {Capetti}, \& {Massaglia}}]{RossiEtal2017A&A...606A..57R}
{Rossi}, P., {Bodo}, G., {Capetti}, A., \& {Massaglia}, S. 2017, \bibinfo{title}{{3D relativistic MHD numerical simulations of X-shaped radio sources},} \aap, 606, A57, \dodoi{10.1051/0004-6361/201730594}

\bibitem[{H. {Rottmann}(2001){Rottmann}}]{Rottmann2001PhDT.......173R}
{Rottmann}, H. 2001, PhD thesis, -

\bibitem[{K. {Rubinur} {et~al.}(2018){Rubinur}, {Das}, \& {Kharb}}]{RubinurEtal2018JApA...39....8R}
{Rubinur}, K., {Das}, M., \& {Kharb}, P. 2018, \bibinfo{title}{{Searching for dual active galactic nuclei},} Journal of Astrophysics and Astronomy, 39, 8, \dodoi{10.1007/s12036-018-9512-y}

\bibitem[{K. {Rubinur} {et~al.}(2019){Rubinur}, {Das}, \& {Kharb}}]{Rubinur2019}
{Rubinur}, K., {Das}, M., \& {Kharb}, P. 2019, \bibinfo{title}{{Searching for dual AGN in galaxies with double-peaked emission line spectra using radio observations},} \mnras, 484, 4933, \dodoi{10.1093/mnras/stz334}

\bibitem[{P.~A.~G. {Scheuer} \& R. {Feiler}(1996){Scheuer} \& {Feiler}}]{ScheuerFeiler1996MNRAS.282..291S}
{Scheuer}, P.~A.~G., \& {Feiler}, R. 1996, \bibinfo{title}{{The realignment of a black hole misaligned with its accretion disc},} \mnras, 282, 291, \dodoi{10.1093/mnras/282.1.291}

\bibitem[{N.~I. {Shakura} \& R.~A. {Sunyaev}(1973){Shakura} \& {Sunyaev}}]{ShakuraSunyaev1973A&A....24..337S}
{Shakura}, N.~I., \& {Sunyaev}, R.~A. 1973, \bibinfo{title}{{Black holes in binary systems. Observational appearance.},} \aap, 24, 337

\bibitem[{S. {Silpa} {et~al.}(2022){Silpa}, {Kharb}, {Harrison}, {Girdhar}, {Mukherjee}, {Mainieri}, \& {Jarvis}}]{SilpaEtal2022MNRAS.513.4208S}
{Silpa}, S., {Kharb}, P., {Harrison}, C.~M., {et~al.} 2022, \bibinfo{title}{{The Quasar Feedback Survey: revealing the interplay of jets, winds, and emission-line gas in type 2 quasars with radio polarization},} \mnras, 513, 4208, \dodoi{10.1093/mnras/stac1044}

\bibitem[{R.~J. {Turner} \& S.~S. {Shabala}(2015){Turner} \& {Shabala}}]{TurnerShabala2015ApJ...806...59T}
{Turner}, R.~J., \& {Shabala}, S.~S. 2015, \bibinfo{title}{{Energetics and Lifetimes of Local Radio Active Galactic Nuclei},} \apj, 806, 59, \dodoi{10.1088/0004-637X/806/1/59}

\bibitem[{S.~D. {von Fellenberg} {et~al.}(2023){von Fellenberg}, {Janssen}, {Davelaar}, {Zaja{\v{c}}ek}, {Britzen}, {Falcke}, {K{\"o}rding}, \& {Ros}}]{vonFellenbergEtAl2023AA...672L...5V}
{von Fellenberg}, S.~D., {Janssen}, M., {Davelaar}, J., {et~al.} 2023, \bibinfo{title}{{Radio jet precession in M 81*},} \aap, 672, L5, \dodoi{10.1051/0004-6361/202245506}

\bibitem[{A. {Wirth} {et~al.}(1982){Wirth}, {Smarr}, \& {Gallagher}}]{WirthEtAl1982AJ.....87..602W}
{Wirth}, A., {Smarr}, L., \& {Gallagher}, J.~S. 1982, \bibinfo{title}{{Dumbbell galaxies and precessing radio jets.},} \aj, 87, 401, \dodoi{10.1086/113135}

\bibitem[{E.~L. {Wright} {et~al.}(2010){Wright}, {Eisenhardt}, {Mainzer}, {Ressler}, {Cutri}, {Jarrett}, {Kirkpatrick}, {Padgett}, {McMillan}, {Skrutskie}, {Stanford}, {Cohen}, {Walker}, {Mather}, {Leisawitz}, {Gautier}, {McLean}, {Benford}, {Lonsdale}, {Blain}, {Mendez}, {Irace}, {Duval}, {Liu}, {Royer}, {Heinrichsen}, {Howard}, {Shannon}, {Kendall}, {Walsh}, {Larsen}, {Cardon}, {Schick}, {Schwalm}, {Abid}, {Fabinsky}, {Naes}, \& {Tsai}}]{WrightWISE..Survey2010AJ....140.1868W}
{Wright}, E.~L., {Eisenhardt}, P. R.~M., {Mainzer}, A.~K., {et~al.} 2010, \bibinfo{title}{{The Wide-field Infrared Survey Explorer (WISE): Mission Description and Initial On-orbit Performance},} \aj, 140, 1868, \dodoi{10.1088/0004-6256/140/6/1868}

\bibitem[{Z.~H. {Xie} {et~al.}(2009){Xie}, {Ma}, {Zhang}, {Du}, {Hao}, {Yi}, \& {Qiao}}]{XieEtAl2009ApJ...707..866X}
{Xie}, Z.~H., {Ma}, L., {Zhang}, X., {et~al.} 2009, \bibinfo{title}{{Estimation of the Viscosity Parameter in Accretion Disks of Blazars},} \apj, 707, 866, \dodoi{10.1088/0004-637X/707/2/866}

\bibitem[{Y. {Yang} {et~al.}(2019){Yang}, {Bartos}, {Gayathri}, {Ford}, {Haiman}, {Klimenko}, {Kocsis}, {M{\'a}rka}, {M{\'a}rka}, {McKernan}, \& {O'Shaughnessy}}]{YangEtal2019PhRvL.123r1101Y}
{Yang}, Y., {Bartos}, I., {Gayathri}, V., {et~al.} 2019, \bibinfo{title}{{Hierarchical Black Hole Mergers in Active Galactic Nuclei},} \prl, 123, 181101, \dodoi{10.1103/PhysRevLett.123.181101}

\end{thebibliography}
\bibliographystyle{aasjournalv7}

\end{document}